\documentclass[usenatbib]{mn2e}
\usepackage{graphicx, amssymb, aas_macros, natbib,array}
\usepackage{mathrsfs}
\usepackage{amsmath}
\usepackage{gensymb}
\usepackage{mathrsfs}
\usepackage{txfonts}
\usepackage{verbatim}
\usepackage[implicit=false,breaklinks,colorlinks,citecolor=blue]{hyperref}

\newcommand{\vmax}{V_{\rm max}}
\newcommand{\vinfall}{V_{\rm infall}}
\newcommand{\rmax}{r^*_{\rm{max}}
	\newcommand{\vpeak}{V_{\rm peak}}}

\newcommand{\mhalf}{M_{1/2}}
\newcommand{\rhalf}{R_{1/2}}

\newcommand{\vrot}{v_{\rm{rot}}}
\newcommand{\mvir}{M_{\rm{vir}}}
\newcommand{\mhalo}{M_{\rm{halo}}}

\newcommand{\rvir}{r_{\rm{vir}}}

\newcommand{\mstar}{M_{\rm \star}}
\newcommand{\msun}{\rm \, M_{\odot}}

\newcommand{\lsun}{L_{\odot}}

\newcommand{\mpc}{\rm \, Mpc}
\newcommand{\kpc}{\rm \, kpc}
\newcommand{\pc}{\rm \, pc}

\newcommand{\kms}{{\rm km \, s}^{-1}}

\newcommand{\lcdm}{$\Lambda$CDM}

\newcommand{\ratio}{\vrot / \sigma}

\newcommand{\mbar}{m^{\rm p}_{\rm bar}}
\newcommand{\mtenq}{\rm m10q_{30}}
\newcommand{\mtenqsat}{\rm m10q_{30} \, Sat}
\newcommand{\mtenv}{\rm m10v_{30}}
\newcommand{\mtenvA}{\rm m10v_{30}}
\newcommand{\mtenvB}{\rm m10v_{30} \, B}
\newcommand{\mtenvC}{\rm m10v_{30} \, C}
\newcommand{\mtenvD}{\rm m10v_{30} \, D}
\newcommand{\mtenvsat}{\rm m10v_{30} \, Sat}
\newcommand{\mtenvF}{\rm m10v_{30} \, F}
\newcommand{\mtenvG}{\rm m10v_{30} \, G}
\newcommand{\mnine}{\rm m09_{30}}
\newcommand{\mnineA}{\rm m09_{30}}
\newcommand{\mnineB}{\rm m09_{30} \, B}
\newcommand{\mtenqlo}{\rm m10q_{250}}

\newcommand{\mtenvloA}{\rm m10v_{250}}
\newcommand{\mtenvloB}{\rm m10v_{250} \, B}
\newcommand{\mninelo}{\rm m09_{250}}
\newcommand{\cmt}{\rm cm^{-3}}
\newcommand{\feh}{\rm [Fe/H]}

\begin{document}
	
	\voffset=-0.6in
	
	\vspace{-0.7cm}
	\title[Be it therefore resolved]
	{Be it therefore resolved: Cosmological Simulations of Dwarf Galaxies with Extreme Resolution\vspace{-0.7cm}}
	\vspace{-0.3cm}
	\author[C. Wheeler et al.]{
		\parbox[t]{0.98\textwidth}{
			Coral Wheeler\thanks{$\!$coral@caltech.edu}$^1$,
			Philip F.~Hopkins$^1$,
			Andrew B. Pace$^{2}$, 
			Shea Garrison-Kimmel$^1$,\\
			Michael Boylan-Kolchin$^{3}$,
			Andrew Wetzel$^{4}$,
			James S.~Bullock$^{5}$,
			Du{\v s}an Kere{\v s}$^{6}$,
			Claude-Andr{\'e} Faucher-Gigu{\`e}re$^{7}$,
			Eliot Quataert$^{8}$
		}
		\vspace*{10pt} \\
		\noindent$\!\!$ $^{1}$TAPIR, Mailcode 350-17, California Institute of Technology, Pasadena, CA 91125, USA, \\		
		\noindent$\!\!$ $^{2}$Mitchell Institute for Fundamental Physics and Astronomy, Department of Physics and Astronomy, Texas A\&M University, 
		College Station, TX 77843, USA\\
		\noindent$\!\!$ $^{3}$Department of Astronomy, The University of Texas at Austin,
		2515 Speedway, Stop C1400, Austin, TX 78712-1205, USA\\
		\noindent$\!\!$ $^{4}$Department of Physics, University of California, Davis, CA 95616, USA \\
		\noindent$\!\!$ $^{5}$Center for Cosmology, Department of Physics and Astronomy, University of California, Irvine, CA 92697, USA \\
		\noindent$\!\!$ $^{6}$Department of Physics, Center for Astrophysics and Space Sciences, University of California at San Diego, 9500 Gilman Drive, La Jolla, CA 92093, USA \\
		\noindent$\!\!$ $^{7}$Department of Physics and Astronomy and CIERA, Northwestern University, 2145 Sheridan Road, Evanston, IL 60208, USA\\
		\noindent$\!\!$ $^{8}$Department of Astronomy and Theoretical Astrophysics Center, University of California Berkeley, Berkeley, CA 94720, USA}
	\vspace{-0.5cm}
	\pagerange{\pageref{firstpage}--\pageref{lastpage}} 
	\pubyear{2018}
	\maketitle
	\vspace{-2.0cm}
	\label{firstpage} 
	
	\begin{abstract} 
We study a suite of extremely high-resolution cosmological FIRE simulations of dwarf galaxies ($M_{\rm halo} \lesssim 10^{10}\msun$), run to $z=0$ with $30\,M_{\odot}$ resolution, sufficient (for the first time) to resolve the internal structure of individual supernovae remnants within the cooling radius. Every halo with $M_{\rm halo} \gtrsim 10^{8.6}\,M_{\odot}$ is populated by a resolved {\em stellar} galaxy, suggesting very low-mass dwarfs may be ubiquitous in the field. Our ultra-faint dwarfs (UFDs; $M_{\ast}<10^{5}\,M_{\odot}$) have their star formation truncated early ($z\gtrsim2$), likely by reionization, while classical dwarfs ($M_{\ast}>10^{5}\,M_{\odot}$) continue forming stars to $z<0.5$. The systems have bursty star formation (SF) histories, forming most of their stars in periods of elevated SF strongly clustered in both space and time. This allows our dwarf with $M_{\ast}/M_{\rm halo} > 10^{-4}$ to form a dark matter core $>200\pc$, while lower-mass UFDs exhibit cusps down to $\lesssim100\pc$, as expected from energetic arguments. Our dwarfs with $M_{\ast}>10^{4}\,M_{\odot}$ have half-mass radii ($\rhalf$) in agreement with Local Group (LG) dwarfs; dynamical mass vs.\ $\rhalf$ and the degree of rotational support also resemble observations. The lowest-mass UFDs are below surface brightness limits of current surveys but are potentially visible in next-generation surveys (e.g. LSST). The stellar metallicities are lower than in LG dwarfs; this may reflect pre-enrichment of the LG by the massive hosts or Pop-III stars. Consistency with lower resolution studies implies that our simulations are numerically robust (for a given physical model).
		
	\end{abstract}
	
	\begin{keywords}
		galaxies: dwarf -- galaxies: formation -- galaxies: star formation -- galaxies: kinematics and dynamics -- Local Group
		\vspace{-0.7cm}
	\end{keywords}
	
	\section{Introduction}
	\label{sec:intro} 
	
	Although the currently favored cosmological paradigm -- $\Lambda$ Cold Dark Matter Theory (\lcdm) -- has been widely successful in predicting the counts, clustering, colors, morphologies, and evolution of galaxies on large scales, as well as a variety of cosmological observables \citep{Eisenstein2005,Viel2008, Reid2010,Komatsu2011}, several challenges have arisen to this model in recent years, most of them occurring at the smallest scales -- those of dwarf galaxies ($M_{\rm \star} \lesssim 10^9 M_{\odot}$; see e.g. \citealt{Bullock2017} for a review.) Among these small-scale challenges, perhaps best known is the Missing Satellites Problem (MSP; \citealt{Klypin1999, Moore1999,Bullock2010b}): counts of galaxies predicted from a naive assignment of stellar mass to dark-matter only simulations of Milky Way (MW)-mass galaxies drastically over-predicts the actual number of currently observed dwarf galaxies around the MW. 
	
	The severity of the missing satellites problem is sensitive to the low-mass edge of galaxy formation: any halo mass threshold below which galaxy formation cannot proceed will result in firm predictions for the abundance and distribution of low-mass galaxies around the Milky Way. In fact, \citet{Graus2018} suggest that the radial distribution of Milky Way satellites requires galaxy formation to persist in dark matter halos with virial temperatures below the atomic cooling limit, potentially presenting an issue in the opposite sense of the classic missing satellites problem. 
	
	Whether there is a well-defined halo mass scale at which galaxy formation ceases to operate, and the precise location of this low-mass cutoff, remain unknown. Any low-mass cutoff almost certainly would be affected by the timing of reionization -- both the onset and the end -- as well as by the overall flux of ionizing photons, the spectrum of the radiation, the proximity to more massive structure and the self-shielding ability of the gas itself \citep{Efstathiou1992, Dijkstra2004, Hoeft2006, Weinmann2007, Onorbe2015}. Likewise, the properties of the lowest-mass galaxies that do manage to form should reflect the imprint of the cosmic reionizing background. So-called ``fossils" of reionization, as first proposed by \citet{Ricotti2005}, are galaxies that managed to form some stars before having their star formation shut down by reionization \citep{Bullock2000}. Observations of six UFDs around the Milky Way show that they have stellar ages that are indistinguishable from ancient globular clusters, with the entirety of their star formation occurring before $z\sim2$ \citep{Brown2014}. This suggests that ultra-faint satellites of the Milky Way are indeed fossils of the reionization era.
	
	However, the fact that all of the UFDs considered in \citet{Brown2014} were satellites, rather than isolated galaxies, makes it more difficult to distinguish this effect from other quenching mechanisms, such as ram-pressure stripping \citep{Gunn1972}. \citet[][hereafter W15]{Wheeler2015} used simulations run with the first generation Feedback in Realistic Environments \citep[FIRE;][]{Hopkins2014a}\footnote{{\href{http://fire.northwestern.edu.}{\url{http://fire.northwestern.edu.}}} FIRE uses the pressure-entropy version of SPH hydrodynamics.} and with baryonic particle masses $\mbar = 250\msun$ to show that both isolated and satellite dwarfs in the UFD mass range had uniformly ancient stellar populations, suggesting that they were indeed reionization fossils. Additionally, \citet{Rodriguez-Wimberly2018} removed halos from dark-matter-only simulations \citep{Garrison-Kimmel2014a} that would have been destroyed by the galactic disk according to their pericentric distance \citep{Garrison-Kimmel2017} and showed that there is a vanishingly small probability that all of the observed ultra-faints fell into the Milky Way by $z=2$, making environmental quenching an unlikely explanation for their ancient stellar ages. 
	
	Another well known and long-standing small-scale challenge to \lcdm~is the Core-Cusp Controversy \citep[CCC;][]{Flores1994, Moore1994, deBlok2010}, in which the predicted density of dark matter halos as measured from dark-matter-only simulations suggests the presence of a central density ``cusp"\footnote{Note that \citet{Baushev2018} argue that the predictions of cusps themselves may be numerical artifacts.} ($\rho\sim r^{-1}$; \citealt{NFW1997}), while observations of some dwarf galaxies suggest that the actual profile shape can flatten at the center, into a shallower-density ``core" ($\rho\sim \rm const$; \citealt{Salucci2000, vandenBosch2000,  deBlok2002, Oh2008}). 
	
	This issue has since been shown to be largely a result of comparing dark-matter-only simulations to observations, and that a dark matter core can be created in galaxies through repeated fluctuations in the gravitational potential via regular expulsions of the galactic gas supply from bursty star formation and its resulting feedback \citep{Mashchenko2006, Governato2012G, Pontzen2012, Chan2015, Onorbe2015, tollet2016}. However, the ability of supernova feedback to reduce the central density of the dark matter halo is limited by the competing effects of the halo potential and the total mass in stars formed \citep{Garrison-Kimmel2013a}. Galaxy formation is highly inefficient in the regime of UFDs ($\mstar \lesssim 10^5$), meaning the supernova energy input per unit binding energy of the dark matter halo is much lower than in higher mass systems. \citet{DiCintio2014a} showed that galaxies with $\mstar/\mvir\lesssim 10^{-4}$ fail to produce enough star formation to significantly alter the inner halo density cusp to a core. Indeed, most cosmological \lcdm~simulations to date fail to form cores in UFDs \citep[][although see \citealt{Read2016} for an idealized study]{Munshi2013, Fitts2017}. Several UFDs have been found to host globular clusters near their centers, leading some authors to argue that their existence and the lack of dynamical friction implied is evidence for cores in these objects \citep{Goerdt2006, Amorisco2017, Caldwell2017}, but there remains little direct observational evidence for cores in UFDs. If stronger evidence arises, it may require new physics in the dark sector (see, e.g., \citealt{Bullock2017, tulin2018, buckley2018} for recent reviews).
	
	Our understanding of the severity of these challenges to \lcdm~is complicated by computational difficulties in dealing with low-mass galaxies. Most cosmological hydrodynamic simulations of dwarfs run to $z=0$ -- including smoothed particle hydrodynamics (SPH), moving-mesh or adaptive-mesh-refinement (AMR) methods -- have a quasi-Lagrangian mass resolution\footnote{Mass resolution is set by a combination of, e.g., the numerical cell/particle masses and the physical minimum gas mass at which a self-gravitating structure can be identified, which are similar in our study but do not have to be in general (see \citealt{Hopkins2017}).} ranging from $\sim 250-10^{4}\msun$. This means that in UFDs, where the mass of the stellar content can be as low as only a few 100s of solar masses, these simulations may be unable to resolve the galaxies; if they do, these ``galaxies" often have very few, and sometimes even a single, star particles \citep{Sawala2014, Wheeler2015, Munshi2017}. This problem becomes more and more challenging at lower masses, because $\mstar/\mvir$ drops rapidly (meaning a larger and larger number of resolution elements are required to represent the small number of stars that should be present at lower $\mvir$). 
	
	As more and more UFDs are discovered and scrutinized at increasing levels of detail, 
	it becomes imperative for cosmological hydrodynamic simulations to continue to push to higher resolution to allow for comparisons with observations. The stellar particle count must not only be large enough to ensure that the galaxy actually forms (i.e. is not a random grouping of particles), but it also must be sufficient to accurately estimate, e.g., the half-light radius for the galaxies, as well as rotation or other higher-order properties. Cosmological simulations must be able to reliably resolve UFDs to make firm predictions for the next generation of telescopes.

	In this paper, we introduce a new set of high-resolution cosmological hydrodynamic zoom-in simulations (GIZMO/FIRE-2) of isolated dwarf galaxies with baryonic particle masses of $\mbar = 30\msun$ -- the highest resolution ever run to $z=0$. This new generation of $\mbar \sim 10 M_{\odot}$ simulations marks a transition point between simulations that treat star formation within a single stellar population in the aggregate and simulations that model the collapse and fragmentation of a molecular cloud into individual stars, and allows us to probe smaller physical scales than previously possible in cosmological simulations. This in turn enables the comparison between a larger set of ``resolved" simulated UFDs and an ever-increasing set of observations at the low-mass end of galaxy formation. We introduce the suite in Section \ref{sec:sims}, give an overview of results and compare to observations in Section \ref{sec:results}, including the star formation histories (SFHs, Section \ref{sec:sfh}), halo structure (Section \ref{sec:struct}), kinematics (Section \ref{sec:kinetic}), and chemical abundances (Section \ref{sec:chem}). We conclude in Section \ref{sec:conclusion}.
	
	\begin{figure*}
		\includegraphics[width=0.99\textwidth]{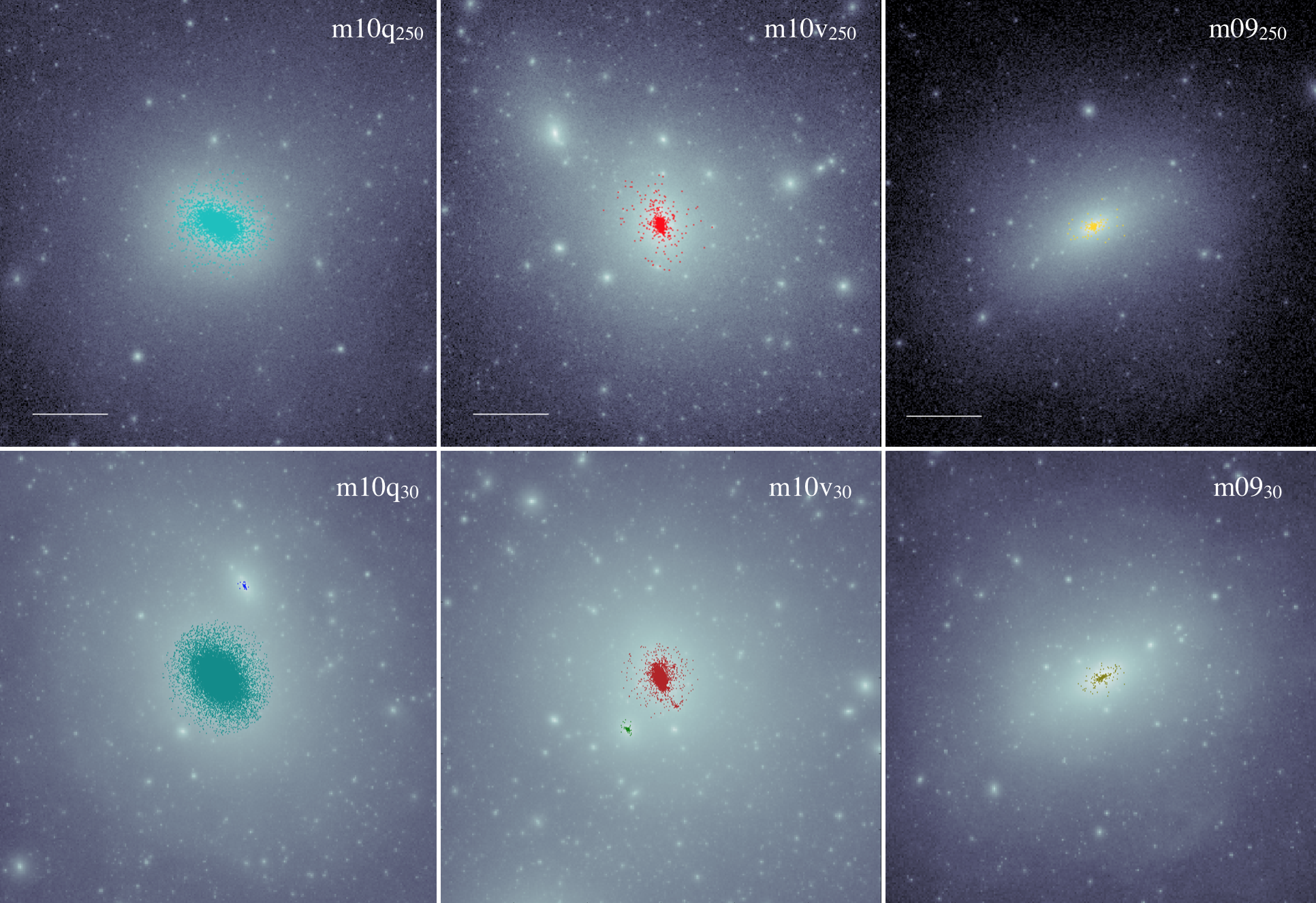} 
		\caption{Dark matter density distribution (greyscale) within the central $30\kpc$ for each simulation in Table~\ref{tab:table}, along with star particles (colors for each simulation) belonging to the central galaxy or a satellite (within $r^*_{\rm max}$; see Table \ref{tab:table}). The top panels show galaxies simulated at ``standard resolution" ($250\,\msun$), while the bottom panels show ``high resolution" simulations ($30\,\msun$). A length scale of $10\kpc$ is shown in each of the top panels; bottom panels are shown at the same scale.
			\label{fig:hists}
		} 
	\end{figure*}
	
	\begin{table*}
		\centering
		\begin{tabular}{| c | c  c  c| c  c  c c | c c c |}
			\hline
			&  $\mhalo (10^{9}\msun)$  &  $\rvir (\kpc)$ & $\vmax \, (\rm kms^{-1})$ &  $\mstar(10^3\msun)$ & $\rmax \, (\rm kpc)$ & $\rhalf \, (\rm pc)$ &  $\ratio$ & $\rm g_{soft} \, (\rm pc)$ & $\rm DM _{soft} \, (\rm pc)$ & $n_{\rm crit}(\cmt)$ \\ 
			\hline
			& \multicolumn{10}{c|}{Resolved Central Galaxies in Our High-Resolution ($30\,M_{\odot}$) Simulations ($>100$ star particles in AHF-Identified Halo)} \\
			\hline
			$\mtenq$ & 7.7 & 51 & 34 & 5200 & 7.7 & 720 & 0.17 & 0.40 & 14 &1000 \\
			$\mtenqsat$ & 0.34 & 6.3 & 16 & 1.2 & 1.2 & 560 & 0.85 & 0.40 & 14 & 1000 \\
			$\mtenvA$ & 9.0 & 54 & 30 & 330 & 8.2 & 330 & 0.45 & 0.10 & 14 & 1e5\\ 
			$\mtenvB$ & 3.2 & 37 &  24  & 41 & 5.6 & 280 & 0.20 & 0.10 & 14 & 1e5\\ 
			$\mtenvC$ & 1.1  & 26 &  16  & 2.9 & 3.9 &  540 & 0.39 & 0.10 & 14 & 1e5\\ 
			$\mtenvD$ & 0.75 & 24 &  16   & 3.7 & 4.8  & 860 & 0.37 & 0.10 & 14 & 1e5\\ 
			$\mtenvsat$ & 0.52 & 20 &  13   & 2.0 & 2.8  &  430 & 0.29 & 0.10 & 14 & 1e5\\ 
			$\mtenvF$  & 0.44 &20 & 12  & 1.9 & 2.5 & 560 & 0.66 & 0.10 & 14 & 1e5 \\ 
			$\mtenvG$ & 0.27 & 17 & 12 & 2.3 & 2.5 & 570 & 1.1 & 0.10 & 14 & 1e5 \\ 
			$\mnineA$ & 2.5  & 35 &  22  & 12 & 4.0 &  200 & 0.56 & 0.10 & 14 & 1e5\\
			$\mnineB$ & 0.67  & 23 &  15  & 1.8 & 3.4  &  620 & 0.79 & 0.10 & 14 & 1e5\\
			\hline
			& \multicolumn{10}{c|}{Resolved Central Galaxies in Our Standard-Resolution ($250\,M_{\odot}$) Simulations ($>100$ star particles in AHF-Identified Halo)} \\
			\hline
			$\mtenqlo$  & 7.5  & 51 &  34   & 2700 & 7.7  &  550 & 0.19 & 1.0 & 29 &1000\\
			$\mtenvloA$ & 8.4 & 53 &  30  & 300 & 8.0 &  310 & 0.31 & 1.0 & 29 & 1000\\ 
			$\mtenvloB$ & 2.7   & 37 &  24  & 66 & 5.5  &  350  & 0.14 & 1.0 & 29 &  1000\\ 
			$\mninelo$ & 2.5  & 36 &  22   & 27 & 5.3  &  420 & 0.14 & 1.0 & 29 &  1000\\
			\hline
			
		\end{tabular}
		\label{tab:table} 
		\caption{Properties of dwarfs in the suite. Each row lists a different {\em resolved, central or satellite} galaxy at $z=0$. Columns give: 
			{\bf (1)} $M_{\rm halo}$: halo mass.
			{\bf (2)} $R_{\rm vir}$: virial  radius. 
			{\bf (3)} $V_{\rm max}$: maximum circular velocity.
			{\bf (4)} $M_{\ast}$: bound stellar mass after removal of satellites  and contamination.
			{\bf (5)} $R_{\rm max}$: radial extent of stars (maximum radius of {\em  any} bound star), as determined from visual inspection.
			{\bf (6)} $R_{1/2}$: mean projected (2D) half-stellar-mass radius.
			{\bf (7)} $v_{\rm rot}/\sigma$: Ratio of the stellar velocity shear $v_{\rm rot}$ to dispersion $\sigma$. 
			{\bf (8)} $g_{\rm soft}$: typical minimum gas gravitational+hydrodynamic force softening reached in star-forming gas (this is adaptive).
			{\bf (9)} ${\rm DM}_{\rm soft}$: dark matter force softening (held constant).
			{\bf  (10)} $n_{\rm crit}$: minimum gas density required for star formation (in addition to self-shielding, Jeans instability, and self-gravity).}
	\end{table*}

	\section{Simulations}
	\label{sec:sims} 
	
	\subsection{Resolution and Motivation}
	Our highest resolution suite consists of cosmological zoom-in simulations of Lagrangian volumes surrounding three ``primary'' isolated dwarf galaxy halos ($\mtenq$, $\mtenv$, $\mnine$), each with baryonic particle mass of $\mbar = 30 \msun$ ($m_{\rm dm}$ is larger by the universal baryon fraction), and $z=0$ virial masses\footnote{We define virial overdensity with the spherical top hat approximation of \citet{Bryan1998}.} $\sim 2-10\times10^{9}\msun$ (see Table \ref{tab:table} for a full list of simulation properties).
	
	Before going forward, we stress that the mass resolution achieved here, $\sim 30\,M_{\odot}$, is not simply an incremental improvement. It reaches a critical physical scale where the cooling radius of a SN remnant (approximately the radius enclosing $\sim 3000\,\msun$, with only very weak residual dependence on metallicity or gas density) is resolved with $\gtrsim 100$ elements, which is essential for capturing the basic dynamics.
	Furthermore, many independent studies (e.g., \citealt{Lapi2005,Kim2015c,Martizzi2015,Walch2015,Hu2018,Hopkins2018b}) have shown that with mass resolution of $\lesssim 100\,M_{\odot}$, predictions for SN feedback become nearly independent of the detailed numerical implementation: whether one simply ``dumps'' $\sim 10^{51}\,{\rm erg}$ into surrounding gas in thermal or energy, or applies a more sophisticated injection model, the asymptotic behavior of the blastwave will converge to the same behavior at this resolution. As a result, \citet{Hopkins2018b} showed the predictions of galaxy-formation simulations become vastly less-sensitive to the sub-grid model for feedback. We will therefore compare various properties at our high resolution and a resolution $\sim  8\times$ poorer. In \citet{Hopkins2017}, the primary galaxies $\mtenq$ and $\mtenv$ are also studied at even lower resolution. 
	
	\vspace{-0.3cm}
	\subsection{Numerical Methods}
	
	All details of the numerical methods and initial conditions\footnote{The ICs used here are publicly available at {\href{http://www.tapir.caltech.edu/~phopkins/publicICs}{\url{http://www.tapir.caltech.edu/~phopkins/publicICs}}}}, are presented in \citet{Hopkins2017}, where lower-resolution versions of these volumes were studied extensively. We briefly summarize essential elements here. 
	
	The simulations are run with the {\small GIZMO} \citep{Hopkins2014b}\footnote{A public version of {\small GIZMO} is available at \href{http://www.tapir.caltech.edu/~phopkins/Site/GIZMO.html}{\url{http://www.tapir.caltech.edu/~phopkins/Site/GIZMO.html}}} code using the updated FIRE-2 implementation of star formation and stellar feedback from \citet{Hopkins2017}. FIRE-2 uses the ``meshless finite mass'' (MFM) Lagrangian finite-volume Godunov method for the hydrodynamics, accounts for gas heating from a variety of processes including the UV background\footnote{We adopt the UV background from the December 2011 update of \citet{Faucher-Guigere2009}  (available here: \href{http://galaxies.northwestern.edu/uvb/}{\url{http://galaxies.northwestern.edu/uvb/}}), which reionizes the universe rapidly around $z \sim 10$ and completes reionization by $z=6$, and was designed to produce a reionization optical depth consistent with WMAP-7} and local sources and cooling from $T=10-10^{10}\,K$, and models star formation via a sink-particle method in gas that is locally self-gravitating, Jeans-unstable, self-shielding/molecular, and exceeds a critical density $n_{\rm crit}$ (see table \ref{tab:table}). We adopt a standard, flat $\Lambda$CDM cosmology with $h \approx 0.70$, $\Omega_{\rm m}=1-\Omega_\Lambda \approx 0.27$, and $\Omega_{\rm b}\approx 0.045$ (consistent with current constraints; see \citealt{Planck2014}).\footnote{For the sake of comparison with other work, $\mnine$ adopts slightly different cosmological parameters than the other two simulations. These differences are at the $\sim 1\%$ level, and matter far less than stochastic run-to-run variance.}

	Once stars form, stellar feedback from SNe (Ia \&\ II), stellar mass-loss (O/B and AGB) and radiation (photo-electric and photo-ionization heating, and radiation pressure, with a five-band radiation transport algorithm; see \citealt{Hopkins2018b}) are included following a {\small STARBURST99} stellar evolution model \citep{Leitherer1999} based on the star particle age and metallicity (adopting a \citealt{Kroupa2002} IMF). In lower-resolution simulations, the feedback quantities (e.g.\ SNe rates, stellar spectra) can simply be IMF-averaged. At the high resolution here, however, this produces un-physical outcomes: for example, it effectively assigns $\sim 1/3$ of an O-star to each new star particle, rather than having $\sim 1/3$ of star particles in young stellar populations contain an O-star. We therefore follow \citet{Ma2015,Su2018,Grudic2018} to sample the IMF discretely. Note that while \citet{Su2018} show in a variety of tests that this has little effect on galaxy properties when compared to run-to-run stochastic variation, \citet{Applebaum2018} show that stochastic sampling of the IMF makes feedback burstier, stronger,
	and quenches star formation earlier in small dwarf galaxies.
	
	\vspace{-0.3cm}
	\subsection{Analysis}
	
	We use the Amiga Halo Finder \citep[AHF;]{Knollmann2009} to identify gravitationally bound members of each halo in post-processing. We initially require $>100$ bound {\em star} particles to consider a galaxy resolved. Because AHF can incorrectly assign central stellar halo stars to subhalos, we visually inspected each simulation to determine whether all star particles within the virial radius of each halo are physically associated with the central galaxy or with a satellite, in some cases adjusting the radial extent, $\rmax$, and stellar mass of the galaxy accordingly. All figures reflect these adjustments. Table \ref{tab:table} summarizes the properties of all dwarfs that meet this criterion. Note that a less conservative $\sim15$-particle cut yields $>100$ additional galaxies; these will be studied in future work. Fig.~\ref{fig:hists} shows the DM and stellar distribution in the primary galaxies of each of our zoom-in simulations as well as in the satellites of $\mtenq$ and $\mtenv$.
	
	\begin{figure*}
		\begin{tabular}{cc}
			\vspace{-0.25cm}
			\hspace{-0.15cm}\includegraphics[width=0.49\textwidth]{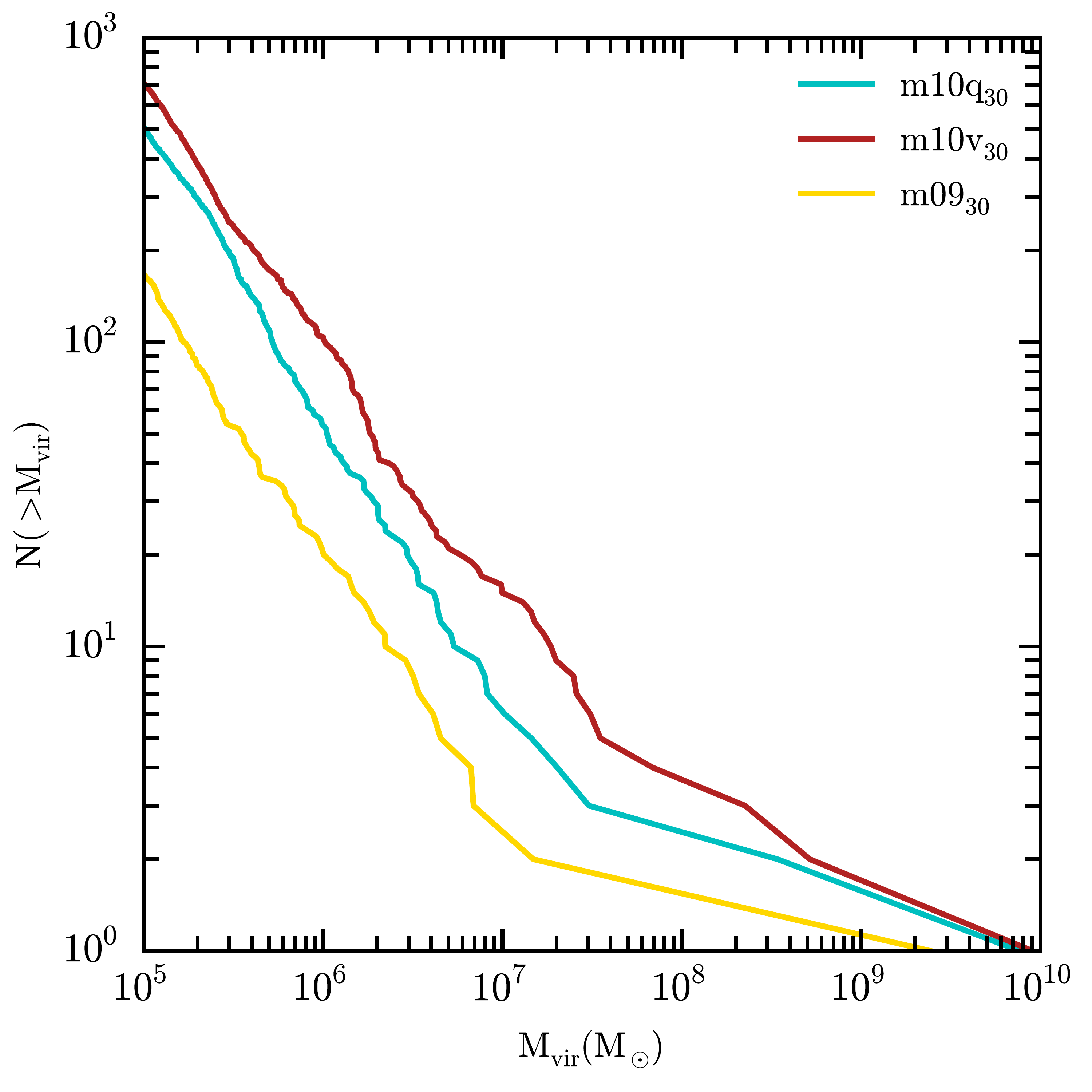} & 
			\hspace{-0.45cm}\includegraphics[width=0.49\textwidth]{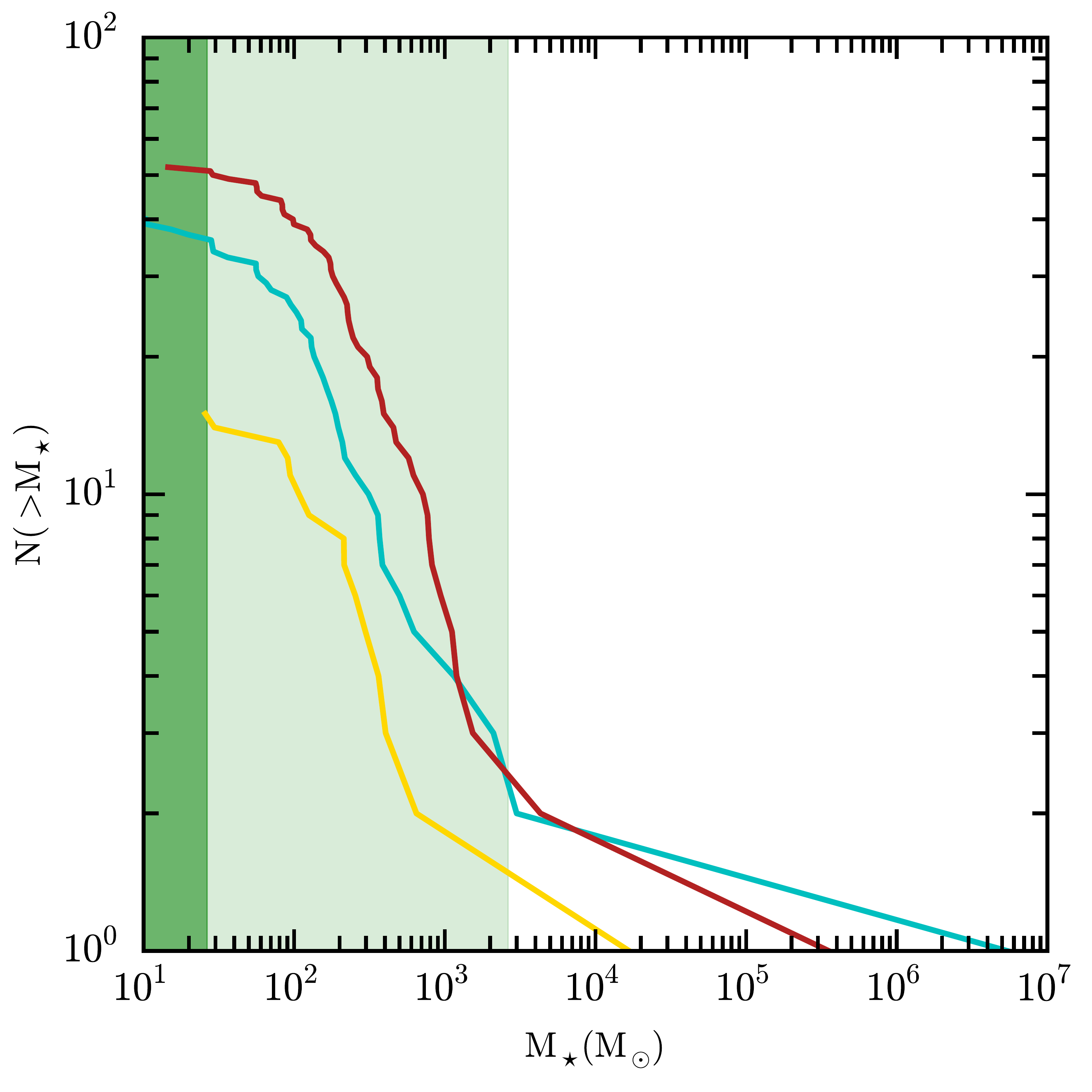} \\
		\end{tabular}
		\vspace{-0.25cm} 
		\caption{{\em Left:} Host+subhalo mass function for our three ``primary" galaxies at $z=0$ (present-day mass is typically a factor $\sim2$ below peak mass). 
			{\em Right:} Stellar mass function for the central + all satellite galaxies that form $\ge1$ star particle. The light (dark) shaded band shows $\le100$ ($\le1$) particle limits for galaxies well-enough resolved for robust estimation of sizes and velocity dispersions.
			\label{fig:massfct}
		} 
	\end{figure*}
	
	\begin{figure}
		\includegraphics[width=0.48 \textwidth]{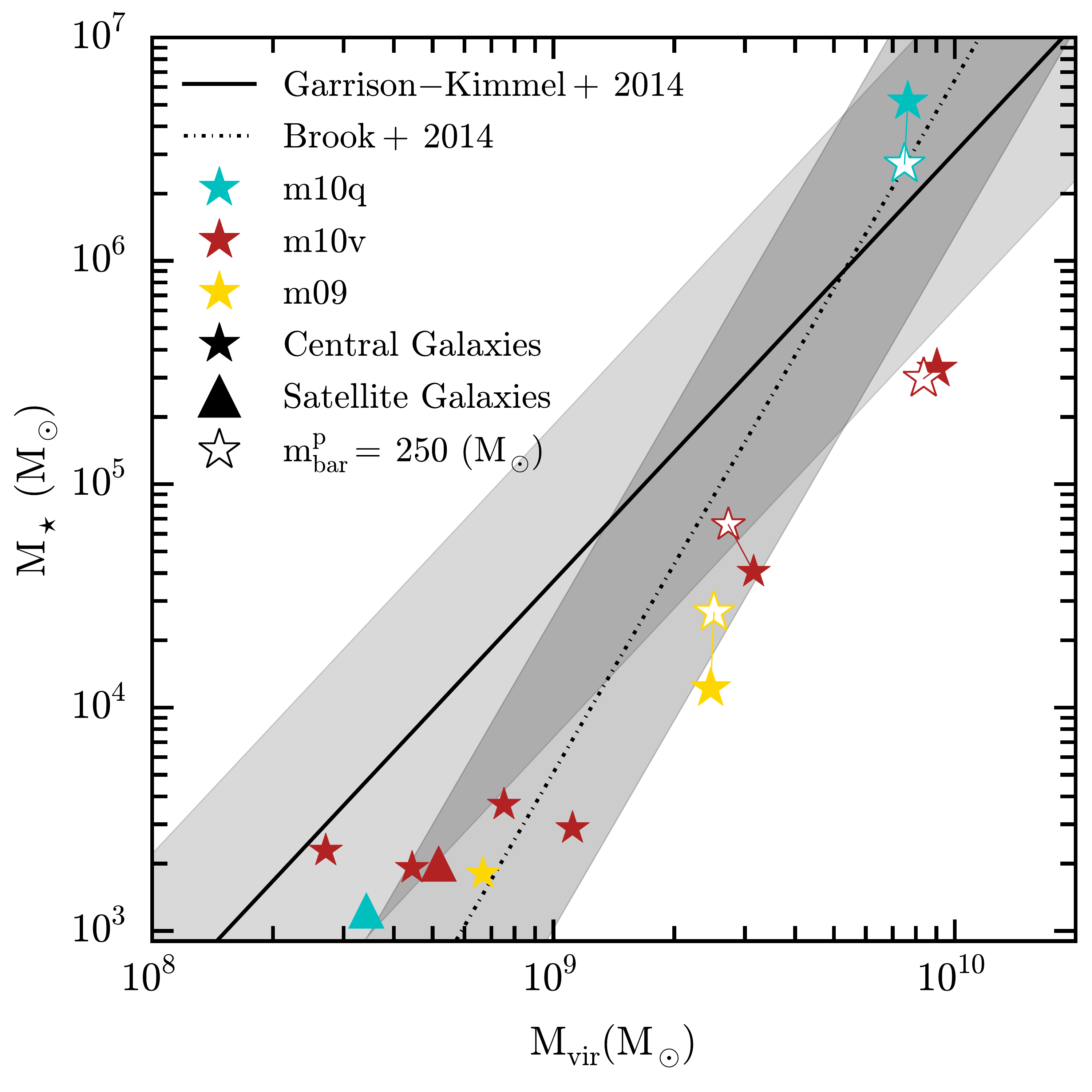} 
		\caption{Stellar mass-halo mass ($\mstar$-$\mhalo$) relation for the resolved central galaxies (Table~\ref{tab:table}; filled/open symbols are the high/low resolution runs) compared to the abundance matching relations from \citet{Garrison-Kimmel2014a} \&\ \citet{Brook2014} with $0.7$-dex scatter. We stress that the comparison is purely heuristic/extrapolated: most of our simulated galaxies are well below detection limits for the observations used to calibrate the relations.
			\label{fig:smhm}
		} 
	\end{figure}

	\begin{figure*}
		\includegraphics[width=1.02\textwidth]{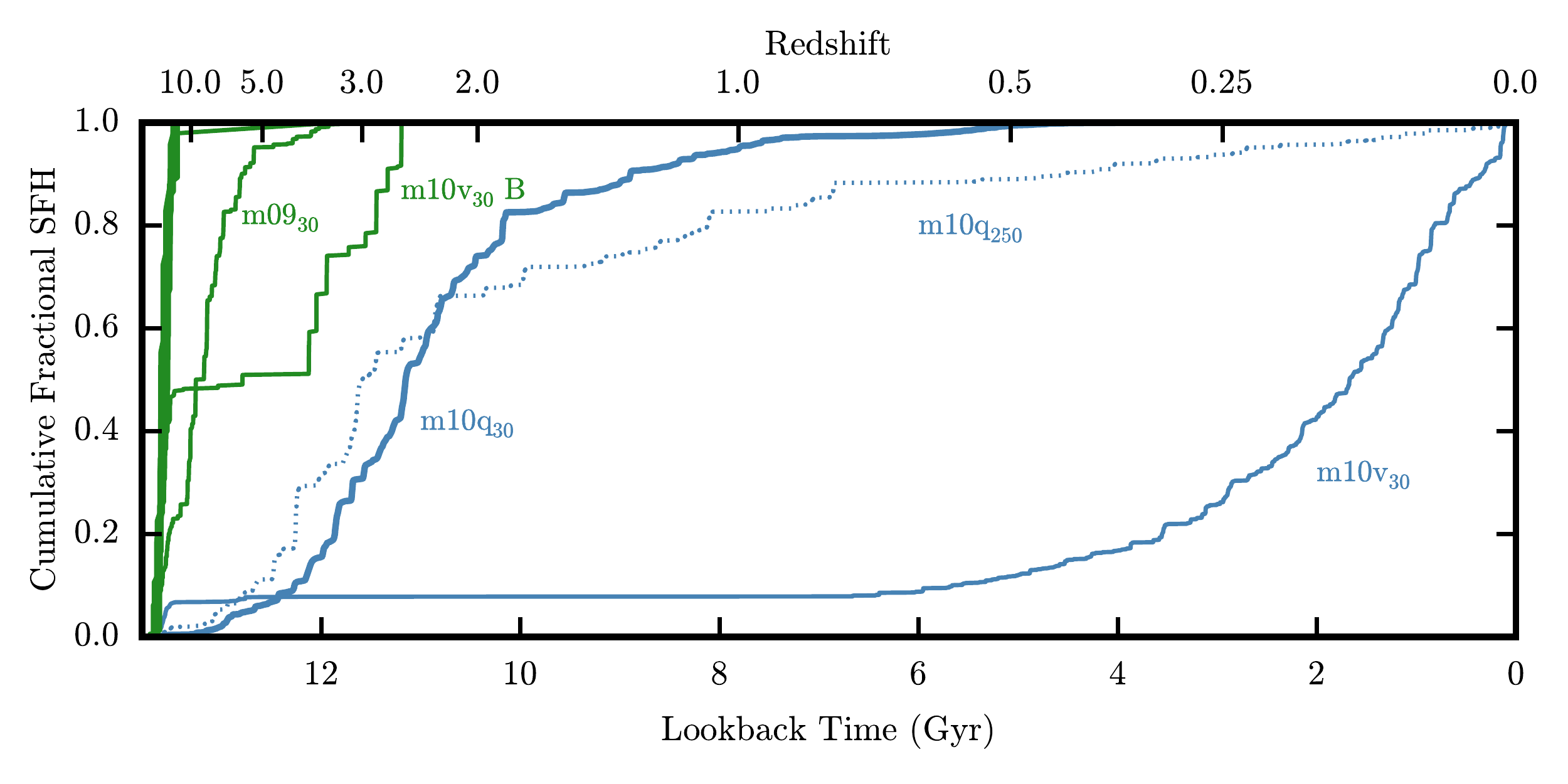} 
		\caption{Star formation histories for each high-resolution galaxy in Table~\ref{tab:table}. Colors distinguish classical dwarfs ($\mstar > 10^5\msun$; {\em  blue}) and UFDs ($\mstar < 10^5\msun$; {\em green}). Every classical dwarf analog formed in our simulations has SF until $z<0.5$, while all UFDs have had their star formation shut down before $z=2$. Because these are isolated UFDs, this suggests that reionization quenched these objects. The dotted line shows $\mtenq$ at lower resolution. Its higher-resolution counterpart (upper blue line) does not show the same trickle of SF at $z=0-0.5$, but this appears to  be stochastic: re-running the low-resolution version with slightly-perturbed initial conditions, we find its SF continues or peters out with approximately equal likelihood. 
			\label{fig:SFH}
		} 
	\end{figure*}
	\begin{figure*}
		\includegraphics[width=1.01 \textwidth]{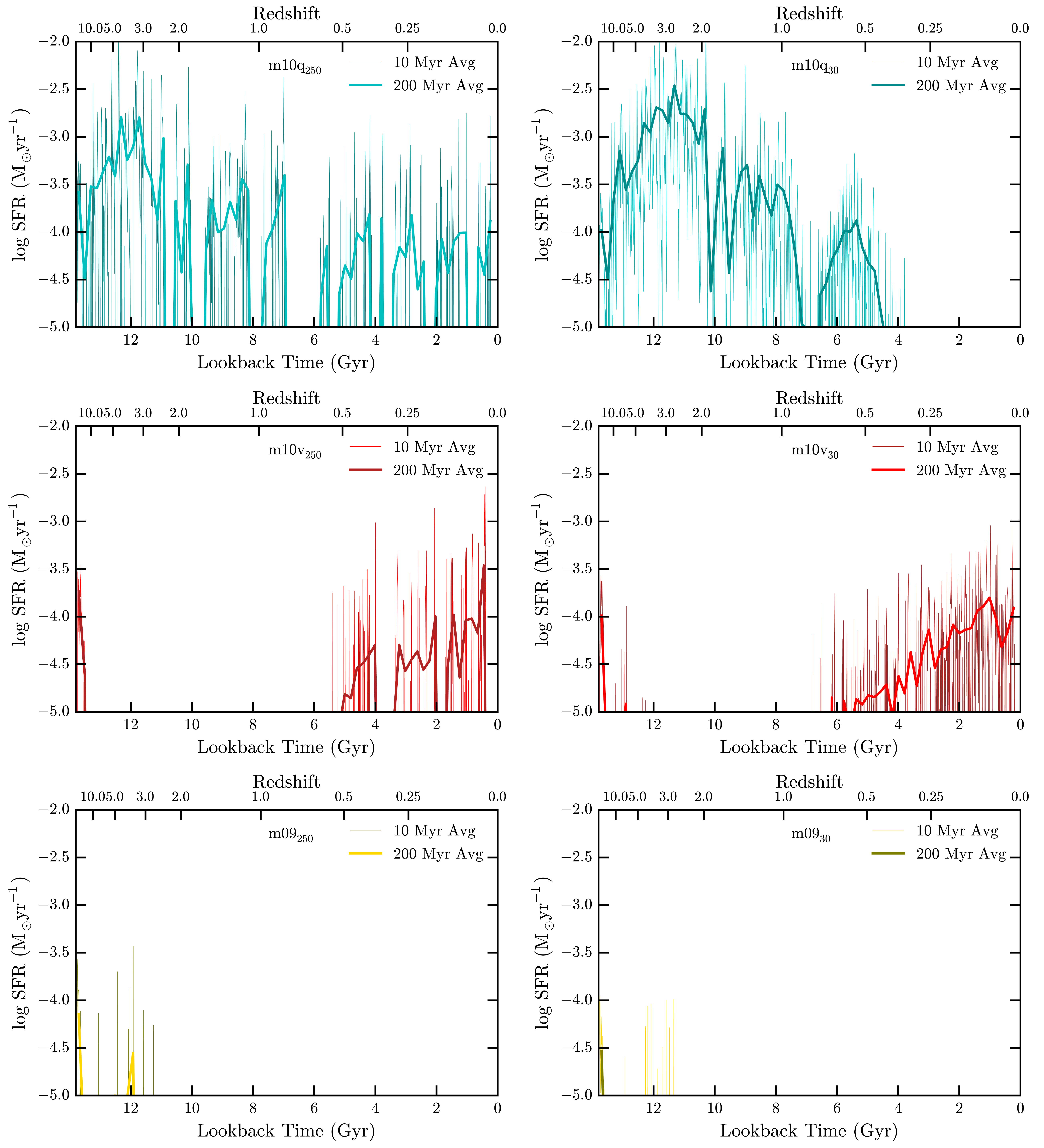} 
		\caption{Star formation rate vs.\ time for the three ``primary'' galaxies resolved in both our lower ($250\,M_{\odot}$; {\em left}) and higher ($30\,M_{\odot}$; {\em right}) resolution boxes. Thin (thick) lines are averaged in $10\,$Myr ($200\,$Myr) intervals. At higher resolution, the gaps where the SFR drops to $\sim 0$ are (at least partially) filled in, although all of the galaxies exhibit considerable burstiness in their SFHs.
			\label{fig:SFR}
		} 
	\end{figure*}

	\section{Results}
	\label{sec:results} 
	
	\subsection{Halo and Stellar Masses}
	
	Fig.~\ref{fig:massfct} shows the DM subhalo and satellite stellar mass functions of the three ``primary" galaxies. Fig.~\ref{fig:smhm} compares the stellar mass-halo mass relation of the simulations for all galaxies in the high resolution region we consider well-resolved to observational constraints implied by abundance-matching.
	
	First, we note that while stochastic run-to-run variations (e.g.\ the magnitude of the largest starbursts at high redshift; see \citealt{ElBadry2017,Su2018,Su2018a,Keller2019}) can lead to changes as large as a factor $\sim 2$ in $M_{\ast}$, there does not appear to be any {\em systematic} dependence on resolution in these properties for the $\sim 4$ galaxies that are well-resolved at both resolution levels. We have also run volumes m10v and m10q at even lower resolution; at resolution  $(30,\,250,\,2\times10^{3},\,1.6\times10^{4},\,1.3\times10^{5})\,M_{\sun}$, the primary galaxy in m10v has stellar mass $M_{\ast}=(3.3,\,3.0,\,1.5,\,4.8,\,2.6)\times10^{5}\,M_{\sun}$, and the primary in m10q has $M_{\ast}=(5.2,\,2.7,\,3.0,\,2.0,1.3)\times10^{6}\,M_{\sun}$. This suggests the sub-grid algorithm that couples SNe mass, energy and momentum to surrounding gas particles properly handles the transition between explicitly resolved and unresolved SNe remnants, as it is specifically designed to do (see \citealt{Hopkins2018b} for extensive tests and discussion). That, in turn, is encouraging for the robustness of previous (lower-resolution) predictions from FIRE simulations.
	
	Our simulated galaxies broadly sample the scatter about extrapolated abundance matching relationships that are tuned to reproduce the stellar mass function of the Local Group, suggesting that the underlying prescriptions in our runs should also reproduce observations when applied to LG-like environments at this resolution. However we stress this comparison is not rigorous: almost all of the simulated galaxies are below the detection limits (in surface brightness or stellar mass) of the observations used to calibrate the abundance-matching relations; moreover, the observations are of LG satellites, not isolated systems. For a more rigorous comparison of classical dwarf mass functions, see \citet{Garrison-Kimmel2018}. 
	
	Our simulations suggest, at least absent a massive host such as the MW, UFDs form in all DM halos with $\mvir^{z=0} \gtrsim 5\times10^8\msun$ (equivalently, $\vmax^{z=0}>13~\kms$). This is just for galaxies with $\gtrsim100$ star particles: we form galaxies with at least 15 star particles in all halos with $\vmax^{z=0}>10~\kms$ and at least 1 star particle in all halos with $\vmax^{z=0}>7.5~\kms$. To get a (very crude) sense of the implications for LG dwarf populations, we can compare to the number of $\vinfall > 13\,\kms$  halos+subhalos (where $\vinfall$ is $\vmax$ when the satellite crosses MW's virial radius) in DM-only simulations of 12 LG analogues from \citet{Garrison-Kimmel2014a}. This naively predicts $\sim 400-610$ such MW satellites with $\mstar \gtrsim 10^{3}\msun$, $\sim180-380$ ``isolated'' dwarfs above this mass in the Local Field\footnote{The Local Field is the immediate environment of the MW and M31 outside of each massive galaxy's virial radii.} (within $1\mpc$), and up to $140$ satellites of other field dwarfs in the same volume, where the range spans one standard deviation from the mean for the sample. 
	
	However, the mere presence of a MW-mass galaxy at the center of a halo has been shown to completely destroy $\sim1/3-1/2$ of all of its subhalos \citep{Garrison-Kimmel2017}, so the number of satellites may be significantly less than these rough estimates. Using a higher $\mstar \gtrsim 10^{4}\,\msun$ ($\vmax\gtrsim 20~\kms$) threshold gives results qualitatively consistent with the abundance-matching inferences for these populations in \citet{Dooley2017}. Interestingly, the fact that we do not see a sharp cut-off in halo mass for UFDs is also consistent with the recent study by \citet{Graus2018}, which argued the radial distribution of MW satellites requires halos with $\vinfall \gtrsim 6-10~\kms$ be populated with UFDs. We will study the less well-resolved, but more abundant, lower-mass UFD population in future work.
	
	\subsection{Star Formation Histories}
	\label{sec:sfh}
	
	Figs.~\ref{fig:SFH}-\ref{fig:SFR} show the archaeological SFH for each galaxy (the distribution of formation times of all $z=0$ stars), in cumulative and differential form. In every case, the more massive galaxies have SF down to $z<0.5$, while all galaxies with $\mstar < 10^5\msun$ had SF cease well before $z=2$. The least massive UFDs, with $\mstar < 10^4\msun$ in halos with $\mhalo \lesssim 10^9\msun$, have the most ancient populations, with SF ceasing before $z=10$ 
	(the midpoint of reionization, as modeled here).\footnote{The one exception appears to be a single star particle from $\mtenvsat$, likely a contaminant from its host $\mtenvA$.} This is consistent with the SFHs of LG dwarfs \citep{Brown2014}. Since our galaxies do not have a massive host, it is clear that the UFD quenching was driven by reionization, not environmental effects. High-mass UFDs ($\mstar \sim 10^{4-5}\,M_{\sun}$) have their accretion cut off by reionization at $z\sim10-6$, but form stars for another $\sim 1$\,Gyr (until $z\sim 2-4$). This residual SF comes from gas that was accreted pre-reionization and is self-shielding to the UV background, so continues to form stars until it is exhausted or blown out by SNe (see also \citealt{Onorbe2015}).

	We note that the primary m10q galaxy appears to ``self-quench,'' i.e. quench without external influence, and exhaust its cold gas before $z=0$. Although this does not occur in the specific lower-resolution run here (which has on-going SF to $z=0$), in several previous studies \citep[see e.g.][]{Su2018} we have shown that this occurs semi-stochastically in lower-resolution runs (as a particularly large burst of SNe concurrently can eject the remaining small amount of cold gas in the galaxy and shut down SF). Roughly, comparing iterations of m10q with intentionally small perturbations to the ICs or run-time parameters in \citet{Su2018}, \citet{Hopkins2017} and the ensemble of dwarfs studied in \citet{Fitts2017}, we find this occurs with order-unity probability. Without a statistical sample, we cannot say if it is more likely in high-resolution runs, but it does indicate that such gas expulsion is not purely an artifact of lower resolution. Whether this is consistent with observations requires a larger sample: \citealt{Geha:2012kx} argue the non-star forming fraction in the field for galaxies with $10^7 \msun<\mstar<10^9\msun$ is zero, but several obviously quenched counter-examples with $\mstar \la 10^7\,\msun$ exist out to $>1$\,Mpc from the LG \citep{Karachentsev2001,Karachentsev2014,Karachentsev2015,Makarov2012,Makarov2017,Cole2014}, and \citet{Fillingham2018} argue that the quenched fraction in the LG and field dwarfs requires some internal quenching (i.e.\ un-connected to a massive host) must occur at these masses.
	
	Fig.~\ref{fig:SFR} shows the SFR averaged in $10$ and $200$\,Myr windows, analogous to timescales over which H$\alpha$ and UV continuum measurements are sensitive; we find the SFRs are bursty, as found previously at lower resolution \citep{Muratov2015,Sparre2017,Faucher-Giguere2018}. SFRs for the other simulated UFDs are similar to $\mnineA$ shown. The results are qualitatively similar at low/high resolution, although the periods of low SFR in the low-resolution runs drop to $0$ while remaining finite at high resolution -- this is simply an artifact of the discrete nature of our star particles (at $\dot{M}_{\ast} \sim 10^{-6}\,\msun\,{\rm yr^{-1}}$, it requires $\sim 250\,$Myr to form a single star particle at $\sim 250\,M_{\sun}$ resolution). Quantitatively, whether the burstiness changes with resolution depends on our definition. The  unit of SF (a star particle) is larger at low resolution, so sampling variations decrease at higher resolution. Based on the metric used in \citet{Hopkins2014a}, namely the dispersion in $\log(\langle \dot{M}_{\ast}(\Delta t_{1}) \rangle/\langle \dot{M}_{\ast}(\Delta t_{0}) \rangle)$ (where $\Delta t_{1} \sim 10\,$Myr and $\Delta t_{0}\sim $\,Gyr), we actually find the burstiness slightly {\em increases} in higher-resolution runs. However, quantifying burstiness in terms of the fraction of the total $M_{\ast}$ formed in periods where $\langle \dot{M}_{\ast}(\Delta t_{1}) \rangle > 1.5\,\langle \dot{M}_{\ast}(\Delta t_{0}) \rangle$ \citep{Sparre2017}, we find slightly decreasing burstiness at higher resolution (for $\Delta t_{1}\sim 10\,$Myr; for much longer averaging times, the effect vanishes). In all cases, the resolution dependence is small ($\sim 10\%$ for order-of-magnitude change in resolution).

	\subsection{Cusps and Cores}
	\label{sec:struct}

	\begin{figure}
		\includegraphics[width=0.5\textwidth]{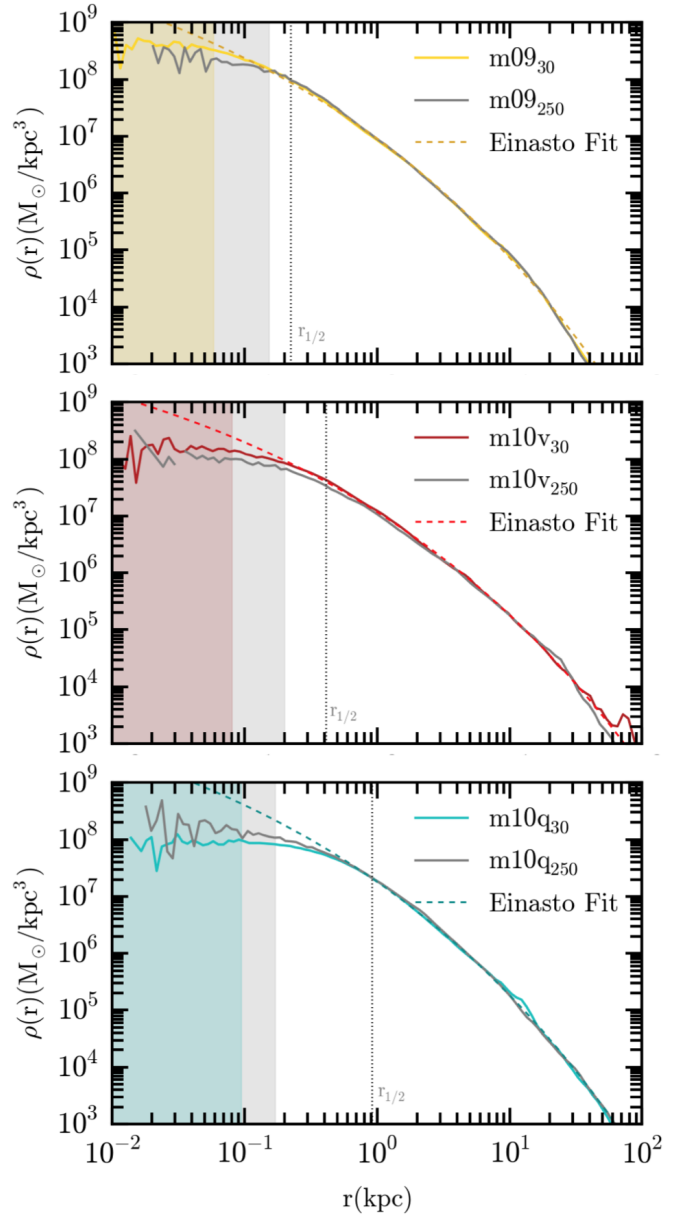} 
		\caption{DM density as a function of radius for the three ``primary'' galaxies from Table~\ref{tab:table}. Results from both standard (grey) and high resolution (colored) simulations are plotted along with an Einasto profile that was fit to the $\mbar = 30$ runs over the range $0.01 < r/\rvir < 1$. The shaded ranges show the regions enclosing $<2000$\, DM particles (colored the same way as the line of the corresponding run). The 3D half-stellar-mass radius is shown as a vertical dotted line.
			The resolution differences are small and do not change the cusp/core distinction. Moreover, the differences are not dominated by dynamical or $N$-body convergence but rather by the differences in the SFHs: runs where the central DM density is slightly higher/lower corresponds to runs which produces slightly fewer/more stars (hence less/more supernovae energy). Although some numerically resolved weak cores are present at $\lesssim 50-100\,$pc, cores with a truly flat log-slope extending to $\gtrsim 300\,$pc only appear when $\mstar/\mvir \gtrsim 10^{-4}$ (only m10q here).
			\label{fig:corecusp}
		} 
	\end{figure}
	
	\begin{figure*}
		\begin{tabular}{cc}
			\vspace{-0.25cm}
			\hspace{-0.15cm}\includegraphics[width=0.49\textwidth]{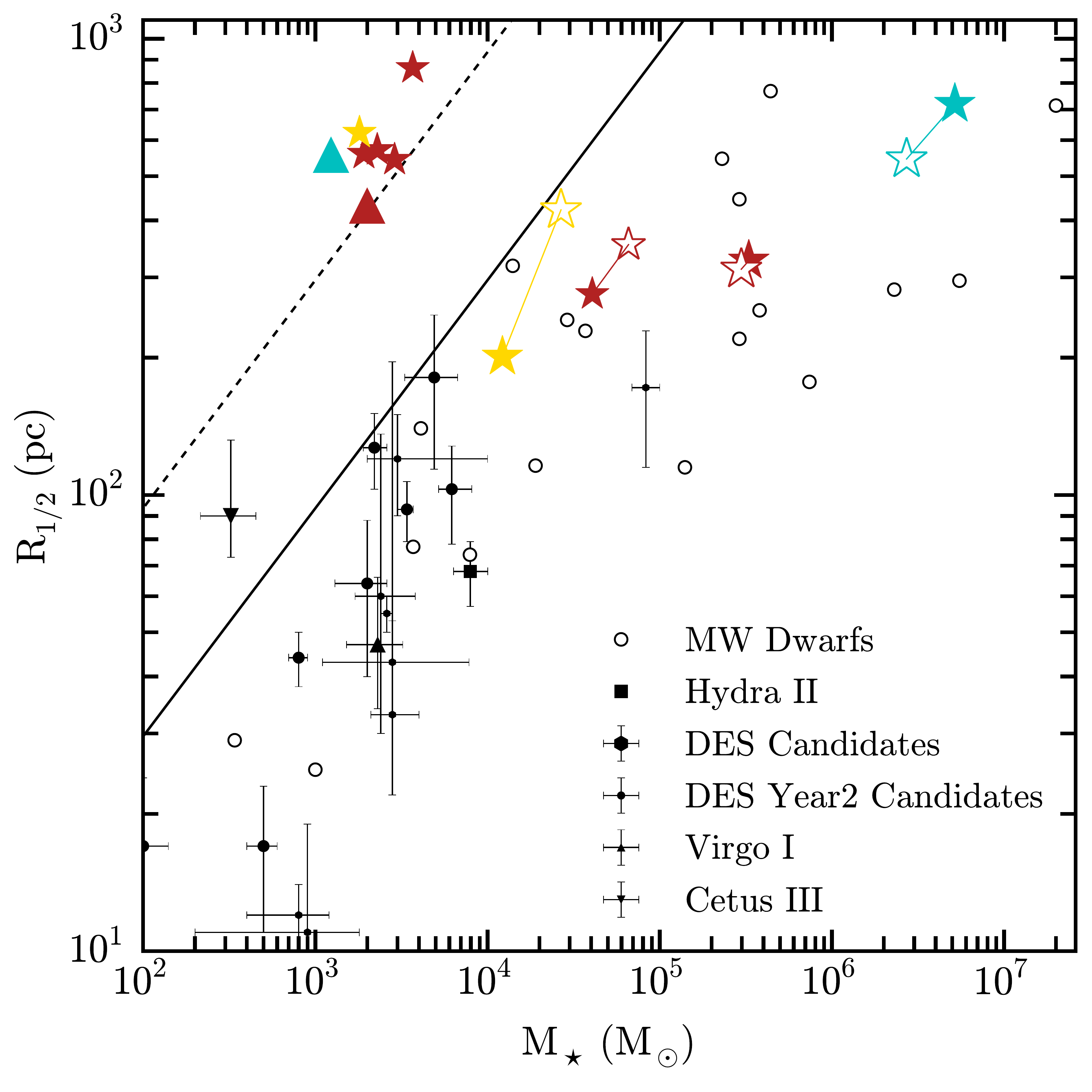} & 
			\hspace{-0.45cm}\vspace{0.5cm}\includegraphics[width=0.49\textwidth]{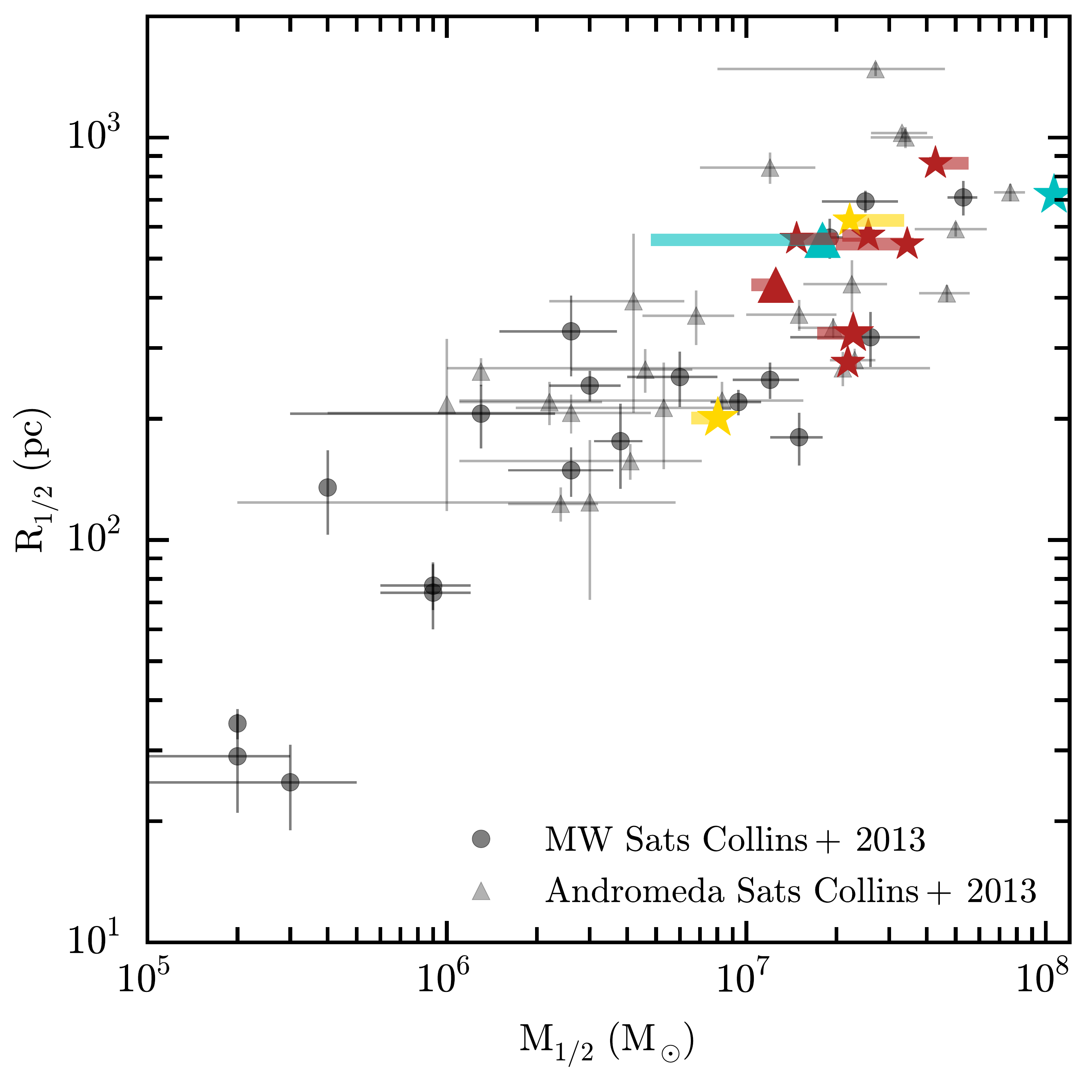} \\
		\end{tabular}
		\vspace{-0.25cm}
		\caption{{\em Left}: 2D half-stellar-mass radii ($R_{1/2}$) vs.\ $\mstar$ for galaxies in Table~\ref{tab:table} (symbols for the simulations as Fig.~\ref{fig:smhm}). We compare our results with observed MW dwarfs compiled from \citep{McConnachie2012}, DES candidates \citep{Bechtol2015}, \citep{Martin2015}, Virgo I \citep{Homma2016}, \&\ Cetus III \citep{Homma2018}.  The solid line is a typical surface brightness limit of these surveys, $30\,\rm mag~arcsec^{-2}$ (for $M/L\approx M_{\odot}/L_{\odot}$). Most of our UFDs will only be visible with future surveys (the dashed line shows a surface brightness limit of $32\,\rm mag~arcsec^{-2}$). 
			{\em Right:} Dynamical mass $M_{\rm half}$ computed using the \citealt{Wolf2010} formula (as is done with observations) vs.\ $R_{1/2}$; a wide colored bar extends from the estimated value to the true dynamical mass inside the 3D half-stellar-mass radius. We also show measured values for MW+Andromeda satellites ($\mhalf$ from \citealt{Collins2013}, $R_{1/2}$ from \citealt{Kirby2014} and \citealt{McConnachie2012}). The agreement in sizes and stellar+dynamical masses is excellent (where observable), except for $\mtenqsat$, which likely suffers from contamination from its host.
			\label{fig:masssize}
		} 
	\end{figure*}

	Fig.~\ref{fig:corecusp} shows the DM profiles of the primary galaxies vs.\ resolution. It is well-known that convergence in the halo mass profiles of  $N$-body calculations is almost entirely determined by mass resolution \citep[see e.g.][]{Power2003}, so the simulations here provide an important convergence test. Using the most conservative definition of a converged radius from \citet{Power2003}, our ultra-high resolution DM profiles should be converged down to $\sim 60-100\,$pc; using the more-aggressive criterion from \citealt{Hopkins2017} gives $\sim 30-40\,$pc. Inside of this 
	Power radius, profiles will tend to flatten for numerical reasons.
	
	However, episodes of strongly-clustered star formation that cycle between dense GMCs and explosive outflow can produce physical cored DM profiles \citep{Pontzen2012,diCintio2014,Chan2015,Onorbe2015}. \citet{Hopkins2017} considered a ``full physics'' resolution study of the DM profiles in m10v and m10q here at lower resolution (from $250$ to $10^{5}\,\msun$), and argued that convergence was dominated by the baryonic effects, not traditional $N$-body considerations, especially in the more massive m10q. We confirm this here: at higher resolution, the core in m10q is {\em more} pronounced, exactly as expected given its {\em larger} stellar mass in that particular run. Given the weak dependence of burstiness on resolution, it is not surprising that the cusp/core behavior also remains robust.
	
	Most importantly, as predicted by the much lower-resolution simulations referenced above, {\em all} our UFDs (which all have $\mstar/\mvir < 10^{-4}$), including those not shown, exhibit cusps down to at least $50-100\,$pc. For those UFDs that exhibit $\sim100\pc$ cores, the deviation from NFW occurs just outside of the converged radius, and is likely in part due to imperfect centering on the dark matter. Furthermore, the half-light radii of the UFDs are all substantially larger than their numerical convergence radii, further indicating that measurable dynamics of such UFDs (which are sensitive to the total mass within the deprojected half-light radius; see below and \citealt{Wolf2010}) would point to dark matter cusps. However, it is important to note that the lack of cores predicted in UFDs is resolution-limited, and allows for the presence of small $\lesssim100\pc$ cores to be detected observationally without posing a challenge to the model. 
	
	\subsection{Kinematics and Galactic Structure}
	\label{sec:kinetic}
	
	\subsubsection{Sizes and Surface Brightness Distributions}
	
	Our result that every isolated dark matter halo with $\mvir > 4.4\times10^8\msun$ forms a UFD suggests that these objects may be ubiquitous in the field. However, these low-mass halos have shallow potential gravitational wells, causing the galaxies that form within them to have larger effective radii and extremely low surface brightnesses \citep[][W15]{Kaufmann2007, Bullock2010, Bovill2011a, Bovill2011b}. This means that, despite their abundance, they may be very difficult to detect. W15 calculated values for $\rhalf$, the 2D projected half-mass radius, for their lowest mass UFDs ($\mstar \lesssim 10^4\msun$) that ranged from $200-500\pc$, making them undetectable with current surveys such as the Sloan Digital Sky Survey \citep[SDSS,][]{SDSS}. However, because the galaxies studied had only tens of star particles, and the size of galaxies is highly sensitive to resolution, there was a possibility that the $\rhalf$ of the UFDs was not resolved. 
	
	The left panel of Fig.~\ref{fig:masssize} shows the mass-size relation for our much better-resolved dwarfs alongside data for classical MW dwarfs (\citealt{McConnachie2012}, open black circles), year 1 and 2 UFD candidates galaxies from the Dark Energy Survey \citep[DES,][]{DES2015, Bechtol2015}, Hydra II \citep{Martin2015}, Virgo I \citep{Homma2016}, and Cetus III \citep{Homma2018}; for the last two galaxies, we used their stellar mass estimates under the assumption of a Kroupa IMF in order to be consistent with the simulations. Also shown in the figure is a surface brightness detection limit of $\rm \mu_V = 30~mag~arcsec^{-2}$ for solar absolute magnitude $\rm M_{\odot,V} = 4.83$ assuming a stellar mass-to-light ratio of $\mstar/L \approx 1\, \msun/\lsun$ (so this corresponds to a physical, bolometric $0.036\,L_{\sun}\,{\rm pc}^{-2}$). For, e.g., a Plummer profile with central surface brightness $\Sigma_{\rm peak} = L/\pi\,R_{1/2}^{2}$, this corresponds to the surface brightness detection limit for SDSS. We also compare the improved limit $\rm 32.5~mag~arcsec^{-2}$ which is anticipated for upcoming surveys such as the co-added Large Synoptic Survey Telescope (LSST). 
	
	Our more massive galaxies agree well with observed systems in Fig.~\ref{fig:masssize}, while the lowest mass UFDs have even lower surface brightnesses than those of W15 in the same mass range. Every UFD with $\mstar < 10^4\msun$ in our sample has $\rhalf > 400$, lying close to or above the likely LSST detection limit, which suggests that they may well go undetected for some time. The sizes of these are much larger than our force softening or the mass-resolution-based \citet{Power2003} convergence radius discussed above ($<30\,$pc), so this is likely robust. Moreover, where we do see (small) changes in size with resolution, the galaxies essentially move along close-to-constant surface brightness tracks.
	
	Interestingly, there does not seem to be any tight correlation in the simulated galaxies between $\rhalf$ and $\mstar$, at $M_{\ast} \ll 10^{7}\,M_{\sun}$. This is yet another indication that galaxy formation prescriptions with simplistic recipes for determining galaxy size, or the assumption that $\rhalf\propto \rvir$ \citep[as in e.g.][]{MMW1998,Kravtsov2013}, fail for the lowest mass galaxies (see also \citealt{Kaufmann2007}). The recent detection of ultra-diffuse galaxies (UDGs) likewise suggests that even more massive dwarfs have wildly varying effective radii (several dex range) for galaxies at a single $M_{\ast}$ \citep{vanDokkum2015}. The {\em apparent} size-mass relation in the observations (Fig.~\ref{fig:masssize}) is clearly an effect of the surface brightness limit (where most of the galaxies pile up, as noted by the authors of the survey studies cited above). Crucially, it is likely that {\em most} UFDs lie at lower surface brightness (larger radii at fixed stellar mass) than we currently can detect. The discovery of a new Milky Way satellite in the Gaia Data Release 2 with $\rm \mu_V = 32.3~mag~arcsec^{-2}$ -- 100 times more diffuse than most UDGs -- is likely the first indication of what is to come in the extremely low surface brightness sky \citep{Torrealba2018}.
	
	It is noteworthy that our simulations do not reproduce those UFDs with half-light radii in the $100\pc$, given how well they agree with many other properties of observed dwarfs. To check whether this is also a selection effect, we restrict those dwarfs in our sample that have average surface brightnesses $\rm >30~mag~arcsec^{-2}$ to an annulus that encloses only the central region of the galaxy with at least that surface brightness. Doing so reduces both $\rhalf$ as well as $\mstar$, which yields $\rhalf\sim100\pc$ ($\rhalf<200\pc$) for $\mnine$ ($\mtenvB$) -- bringing both into the region occupied by the DES UFDs. This suggests that observations may only be sensitive to the ``bright'' core of more massive objects and raises an intriguing question: do the DES UFDs all have diffuse (and relatively massive) outer halos that are currently invisible to us?

	\begin{figure*}
		\begin{tabular}{cc}
			\vspace{-0.25cm}
			\hspace{-0.15cm}\includegraphics[width=0.49\textwidth]{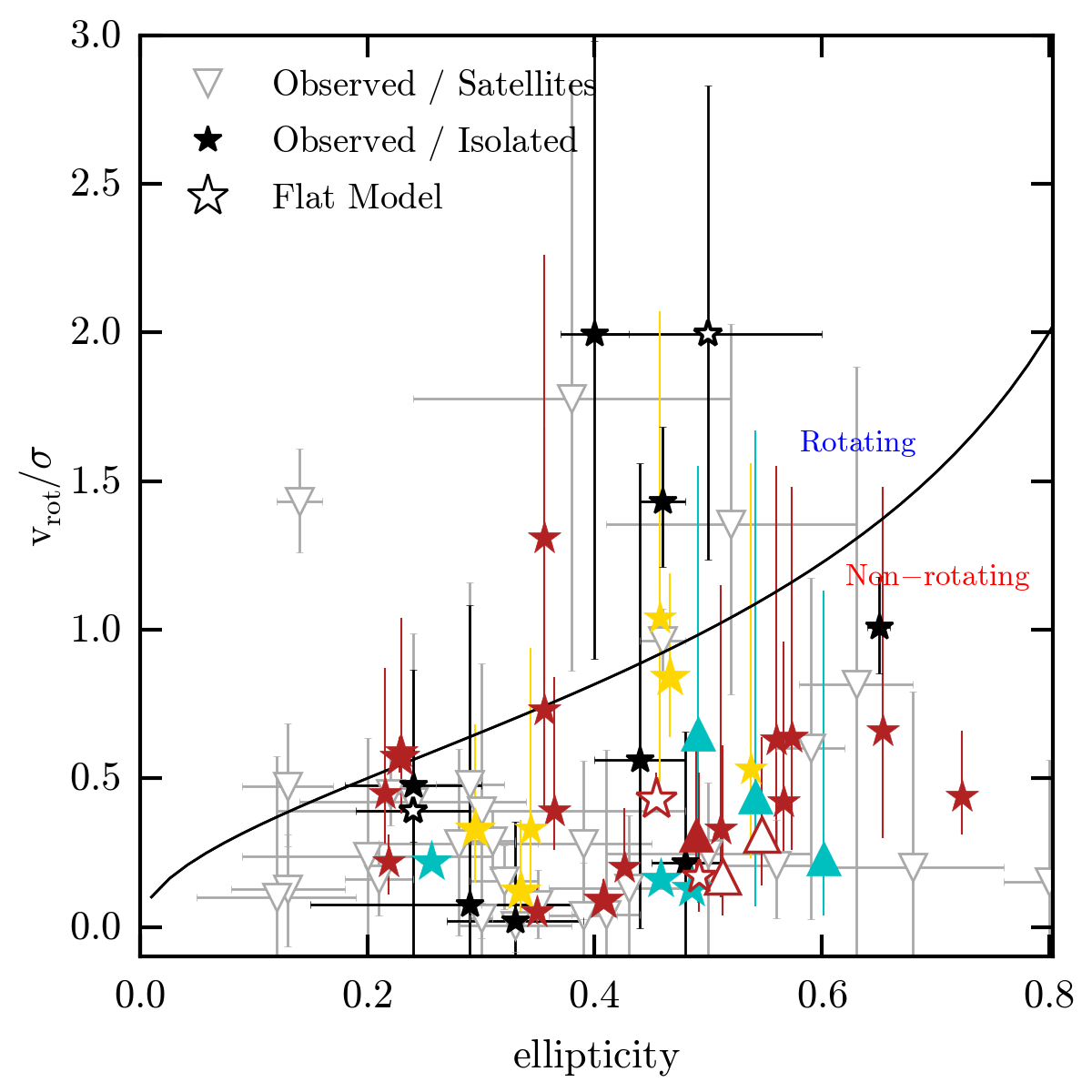} & 
			\hspace{-0.45cm}\includegraphics[width=0.49\textwidth]{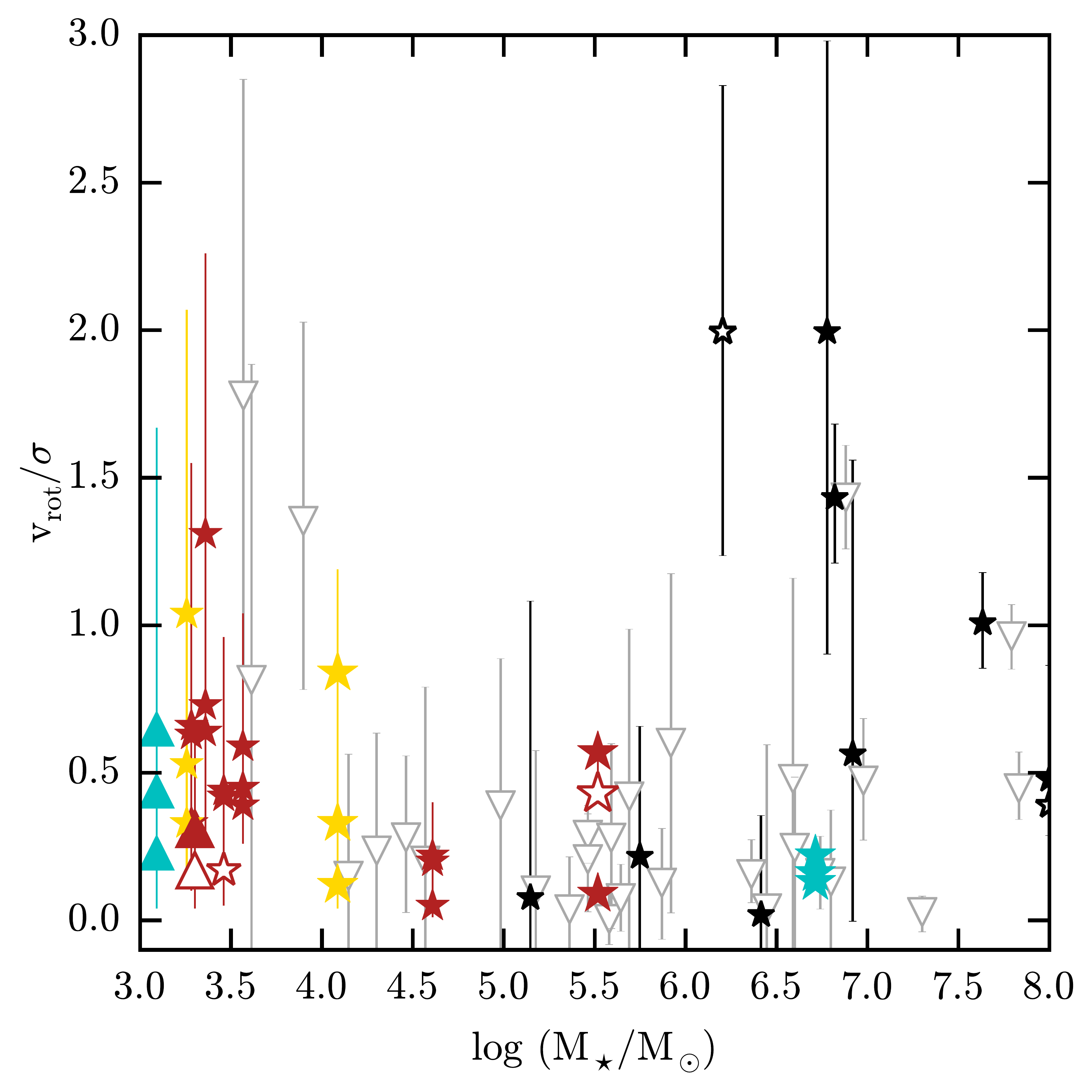} \\
		\end{tabular}
		\vspace{-0.25cm}
		\caption{{\em Left}: Stellar rotation support $\ratio$ vs.\ ellipticity for observed isolated Local Group (LG) dwarfs (compiled in \citealt{Wheeler2017}) and our simulations; the distributions are very similar. Open symbols for the simulations and the isolated observations prefer a flat rotation model. None of the observed satellites prefer a flat model.
			The solid line shows the expectation for self-gravitating systems flattened entirely by rotation (as opposed to anisotropy; \citealt{Binney1978}); most UFDs (field or satellite) are not rotation-dominated. 
			{\em Right:} $\ratio$ vs.\ stellar mass. The distributions are again very similar where they overlap in mass. At high mass ($>10^{6}\,M_{\odot}$), we only have one galaxy (m10q, which has low but significantly non-zero $v_{\rm rot}/\sigma$), so more statistics are needed.
			\label{fig:vsig}
		} 
	\end{figure*}
	
	\vspace{-0.3cm}
	\subsubsection{Dynamical Masses and Sizes}

	Although our lowest-mass simulated galaxies show no overlap with observations in the stellar mass -- size plane, there \textit{is} significant overlap with observations in the dynamical mass -- size plane.  The right panel of Fig.~\ref{fig:masssize} shows $\rhalf$ vs.$\mhalf$, where $\mhalf \simeq4\,\langle \sigma_{\rm los}^{2}\rangle\,\rhalf/G$ $\mhalf$ is an estimate of the total (dynamical) mass from $\sigma_{\rm los}$, the line-of-sight stellar velocity dispersion \citep[see, e.g.,][]{Wolf2010}. We compare observed systems from \citet{Collins2013} with $\rhalf$ taken from \citet{Kirby2014} where available, and \citet{McConnachie2012} otherwise. The simulations agree well with the data at the radii/masses both sample (see also \citealt{Campbell2017,Gonzalez2017,Errani2018}), and with the true mass within the 3D half-stellar-mass-radius (endpoints of colored bars).
	
	The overlap between all simulated and observed galaxies in the right-hand panel of Fig.~\ref{fig:masssize} means that the very diffuse galaxies in the left-hand panel, which would be invisible to current observational surveys, do not inhabit dark halos that are fundamentally different from those hosting analogs of observed systems. The truly remarkable aspect of these galaxies is their large half-light radii; the underlying mass distributions, which are strongly dominated by dark matter at all radii, look just like standard predictions of CDM. For $r \ll r_{\rm s}$, where $r_{\rm s}$ is the NFW scale radius, we expect that $M_{\rm enclosed}(<\!r) \propto r^{2}$; this is precisely the correlation seen in Figure~\ref{fig:masssize}.

	\vspace{-0.3cm}
	\subsubsection{Rotation}
	\label{sec:rot}
	
	Observations of both dSphs and dIrrs in the Local Volume suggest that most dwarf galaxies have kinematics dominated by random motions \citep[][hereafter W17]{Wheeler2017}. This can be seen in Fig. \ref{fig:vsig} for an observational sample of thirty dSphs and dwarf ellipticals and ten isolated dwarfs in the LG taken from W17. Following their prescription to calculate $\ratio$ for our simulations, where $\vrot$ is the rotation across its axis and $\sigma$ is the underlying (constant) line-of-sight velocity dispersion, we perform a Bayesian analysis on the positions and velocities of each simulated galaxy along independent lines of sight. We explore two models for $\vrot$: a flat model that assumes constant rotation, $\vrot(R) = v_o$, and a radially varying pseudo-isothermal sphere, $\vrot = v_o \sqrt{1 - R_o/R \arctan(R/R_o)}$, where $R$ is the distance from the rotation axis on the plane of the sky and $v_o$ and $R_o$ are the rotation velocity and rotation radial scale parameters respectively (see W17 for details).
	
	Fig.~\ref{fig:vsig} shows $\ratio$ vs ellipticity for all simulations (with three independent viewing angles for each), as well as data from W17. The solid line is the locus occupied by oblate isotropic rotators (galaxies flattened primarily by rotation; see \citealt{Binney1978}). Objects below the line are generally understood to be pressure-supported. Seven of the ten isolated dwarfs (black stars) show no clear signs of rotation; the same is true for nearly all of the LG satellites (grey downward triangles). In addition to posterior probabilities for $\vrot$ and $\sigma$, the model calculates the Bayesian evidence for both rotation vs non-rotation and flat model vs. radially varying models. Objects that prefer a flat model are shown as open symbols for the isolated observed dwarfs and the simulations. None of the observed satellites prefer a flat model. Fig.~\ref{fig:vsig} also shows $\ratio$ vs.\ $M_{\ast}$. 
	
	Our simulated galaxies show a remarkably good overlap with the observed data in both of these planes. With higher resolution and the newer FIRE-2 code, we find marginally more rotation than the FIRE-1 simulations from W17, but the difference is small. Although all three of the more massive dwarfs do have Bayesian evidence for rotation, none have high $\ratio$. At high masses ($\gtrsim 10^{6}\,M_{\sun}$), our sample has just one galaxy (m10q) that has low $\ratio$. Since most observed systems are also non-rotating at this mass, this is expected, but large $\ratio\sim2$ begins to appear at these masses; it therefore would be interesting to explore this mass regime with better statistics.
	
	\subsection{Chemical Abundances}
	\label{sec:chem}
	
\begin{figure}
		\vspace{-0.25cm}
		\hspace{-0.15cm}\includegraphics[width=0.50\textwidth]{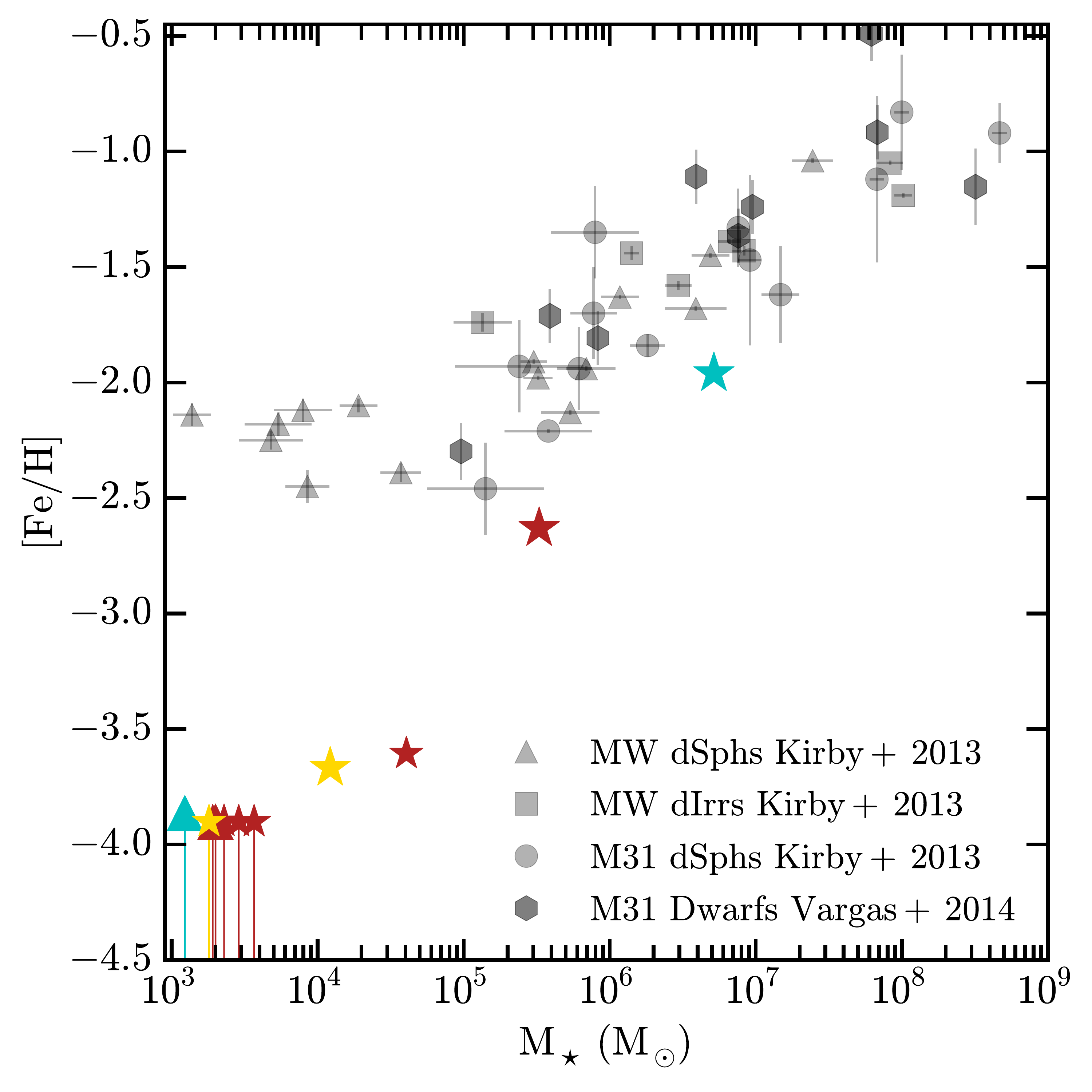} 
		\vspace{-0.25cm}
		\caption{Stellar mass ($\mstar$) vs.\ metallicity ($\feh$) for LG dwarfs (\citealt{Kirby2013}, grey points) and our simulations (colored symbols). The simulated UFDs at $\feh \sim -4$ are at the baseline initial metallicity of the simulation (hence arrows indicating upper limits). While more massive ($\gg 10^{7}\,M_{\sun}$) galaxies agree very well with the observed mass-metallicity relation (see \citealt{Ma2017a,Ma2017b,Escala2018}), below $\sim 10^{7}\,M_{\sun}$ the simulations begin to predict lower $\feh$, with the discrepancy increasing towards lower masses. This may indicate satellites in the LG were pre-enriched by the (much more massive) MW+Andromeda, or by Pop-III stars (not included here; see text for discussion).
			\label{fig:chem}
		} 
	\end{figure}

	\citet{Kirby2013} used stellar metallicities to demonstrate that the relationship between stellar mass and stellar metallicity extends, unbroken, down to Milky Way and M31 dwarf spheroidals, irregulars, and even UFDs down to $\mstar \sim10^{3.5}\msun$. This is a striking result, as the sample includes both satellites and isolated dwarfs, meaning the relationship is unaffected by infall into a more massive host. Fig.~\ref{fig:chem} explores the extremely low-mass end of the (stellar) mass - (stellar) metallicity relation (MZR), comparing observations from \citet{Kirby2013} and \citet{Vargas2014} for MW+Andromeda satellites. There is an obvious and intriguing discrepancy between observed and simulated galaxies, which increases to nearly 2 dex towards lower masses.
	
	We emphasize that at higher masses, $\gg 10^{7}\,M_{\odot}$, previous studies with the same FIRE physics have extensively compared stellar and gas-phase MZRs at both $z=0$ and higher redshifts and found remarkably good agreement between observations and simulations in $\feh$ as well as other species ([O/H], [Mg/H], [Z/H], etc.; see \citealt{Ma2017a,Ma2017b,Escala2018,Hopkins2017,ElBadry2017,Wetzel2016}). The discrepancy reported here appears to be specific to low-mass dwarfs. We also emphasize that the discrepancy is not unique to $\feh$, as  [Mg/H] and other species show a similar (albeit slightly weaker) offset, nor to the method by which $\feh$ is weighted (e.g., light-or-mass weighting or taking $\langle$[Fe/H]$\rangle$ vs.\ [$\langle$Fe$\rangle$/$\langle$H$\rangle$] give a qualitatively similar result).
	
	What could cause this? Recall, our simulations adopt a very simple IMF-sampling and yield model: even though SNe are discrete, the yields are IMF-averaged, and most yields are independent of the progenitor metallicity (see \citealt{Hopkins2017} for details); they therefore reflect yields at solar metallicity, where they are best understood both theoretically and observationally. Yield predictions for extremely metal-poor stars in the literature differ by as much as $\sim 1$\,dex \citep{Woosley1995,Francois2004}, so a strong dependence on progenitor metallicity could explain the offset seen here. It is also possible that the discrepancy has its origin in differential re-incorporation of metals (vs.\ ejection in galactic winds): our dwarfs {\em produce} sufficient metals to lie on the MZR, if they retained them all and re-incorporated them into new stars in a closed-box fashion. But this is difficult to reconcile with the strong outflows required to explain their very low stellar masses \citep{Kirby2011}.
	
	Perhaps more importantly, our simulations include no explicit model for hyper metal-poor or metal-free Population-III stars (this is the reason why they are initialized with a metallicity ``floor'' of $\feh=-4$). Theoretical calculations  suggest a {\em single} massive Pop-III star exploding in a pair-instability supernova could produce $\gtrsim  100\,M_{\sun}$ in heavy elements \citep{Kozyreva2014} -- sufficient to enrich $\gtrsim 3\times10^{6}\,M_{\sun}$ worth of gas (comparable to the mass of our {\em most massive} galaxy studied here) to the minimum metallicity $\feh\sim -2.5$ observed. Indeed, a number of recent studies have argued that an early Pop-III phase should pre-enrich almost all star-forming galaxies in halos more massive than $10^{8}\,M_{\sun}$ to $\feh\sim-2$, even by $z\gtrsim 15$ \citep{Chen2015,Jaacks2018}. Since these would leave no other relics, they would not change our other predictions, but presumably they would leave unique, testable abundance patterns in the observed dwarfs.
	
	Another interesting possibility is that pre-enrichment occurs not via Pop-III stars but environmentally (from the more massive MW/Andromeda host), as the observed dwarfs are all LG satellites. Given typical progenitor masses of MW+Andromeda-mass systems at $z\gtrsim 6$, if just $\sim 1\%$ of the metals produced by SNe at these times escapes to $\sim 100\,$kpc physical radii, then a $\sim 1\,$Mpc co-moving volume would be polluted to $\feh\sim-2$. To test this, we consider the simulations (using the identical code and physics) of LG-mass, MW+Andromeda-like pairs, presented in \citet{Garrison-Kimmel2018}. While the resolution of these simulations ($\sim 3000\,M_{\sun}$) is extremely high for such massive halos, it is not sufficient to resolve the lowest-mass UFDs. However, we can directly compare $\feh$ values for galaxies at $\mstar > 10^5\msun$. We find that there is still a $0.5-1$ dex offset between the simulated satellite galaxies within $300\kpc$ of the galactic center and the observations, suggesting that pre-enrichment from a massive neighbor is not enough to relieve the discrepancy for massive dwarfs. 
	
	To get a sense for how the presence of a host galaxy may affect lower-mass satellites, we consider that all of the galaxies with $\mstar<10^5\msun$ from \citet{Kirby2013A} are within $150\kpc$ of the Galactic center. If we measure the $\feh$ values for the gas enclosed within the radius occupied by $90\%$ of those stars traced back to higher redshift, while excluding the inner $10\%$ to eliminate the galactic disk, we determine that $\sim20\%$ ($1\%$) of the gas has been enriched to $\feh\sim-2.5$ by $z=4$ ($z=10$). This means that, for the more massive UFDs that are still forming stars to $z\sim2$, pre-enrichment from nearby massive host could explain the discrepancy. Additionally, preliminary examination of an extremely-high resolution ($\sim 900\,M_{\sun}$) version of the MW-mass halo studied in \citet{Wetzel2016}, run to $z\sim 4$, shows several satellites pre-enriched to $\feh \sim -2.5$ by $z=5$. However, for the lowest mass UFDs that have completed star formation by $z=10$, this mechanism alone is insufficient to explain the discrepancy.

	\section{Conclusions}
	\label{sec:conclusion}
	
	We have run a suite of hydrodynamic cosmological zoom-in simulations with explicit treatment of star formation and stellar feedback from SNe, stellar mass-loss, and radiation to $z=0$. These simulations have baryonic mass resolution of $30\,M_{\sun}$ (maximum resolved cold gas densities $\gtrsim 10^{5-6}\,M_{\sun}$, spatial scales $\sim 0.1-0.4$\,pc, timescales $\sim 10-100$\,yr), allowing us to probe smaller physical scales than previously possible in cosmological simulations. They resolve ultra-faint dwarf galaxies ($\mstar < 10^5\msun$) with $\gtrsim 100-1000$ star particles and, for the first time, the internal structure of of individual SNe remnants within the cooling radius. Using these simulations, we have shown that:
	
	\begin{itemize}
		\item ``Well-resolved" galaxies ($\gtrsim 100$ star particles) exist in all DM halos with $\mhalo > 5\times10^8\msun$ or $\vmax > 13~\kms$. This suggests these are ubiquitous in the field. Using DM-only LG simulations, we estimate there are anywhere from $\sim180-380$ such UFDs ($\mstar \sim 10^{3-5}\msun$) ``isolated'' in the Local Field, up to another $\sim 140$ as satellites of other Local Field dwarfs,  and $\sim 400-610$ within the virial radii of the MW and Andromeda. However, we also predict most UFDs have very low surface brightness, below the detection capabilities of current surveys. It may be more efficient to search for UFDs as satellites of isolated LG field dwarfs: in the simulations, these have properties that may render them detectable by fully co-added LSST data.
		
		\item All UFDs have uniformly ancient stellar populations: they complete star formation by $z>2$, even if they are isolated field centrals, owing to reionization quenching. More massive field dwarfs continue SF to $z<0.5$.
		
		\item Dwarfs in this mass range have bursty and strongly clustered SF in both time and space, which produces violent outflows. Despite this, only dwarfs with $\mstar/\mvir \gtrsim 10^{-4}$ produce appreciable cores in their DM profiles; all the UFDs are below this threshold and exhibit DM cusps down to $\sim 100\,$pc or smaller (equivalent to $\sim0.2\,R_{1/2}$). This is expected from simple energetic arguments: the UFDs simply produce insufficient stellar feedback to strongly perturb their DM potential. Neither the burstiness nor cusp/core threshold is strongly sensitive to resolution.
		
		\item Properties that can be measured in lower-resolution simulations (e.g., masses, sizes, cusp/core profiles, SFHs of the more massive galaxies here) appear robust to resolution, given the FIRE-2 numerical method adopted here. This is particularly important since many studies have argued the evolution of SNe blastwaves, bubble overlap, and galactic outflows becomes much less strongly-sensitive to sub-grid numerical implementation choices at a mass resolution $< 100\,M_{\sun}$ (owing to the ability to resolve individual SN blast waves self-consistently). 
		
		\item Where the surface brightness is high enough to be detected, the simulations agree well with the location of observed galaxies in size-stellar mass-dynamical mass space. While there is a physical correlation between stellar size ($R_{1/2}$) and dynamical mass, driven by sampling a larger fraction of the halo mass at larger radii, there is no {\em intrinsic} tight correlation between stellar mass and size at these UFD masses. Rather, the observed relation reflects the surface-brightness limits of current surveys.
		
		\item  The simulations have $\ratio$ and ellipticity values consistent with most observed LG isolated and observed dwarfs. Both simulated and observed systems are primarily non-rotating, even in star-forming, isolated field galaxies. Hints of discrepancies between rotation in observed and simulated FIRE dwarfs suggested in \citet{ElBadry2017b,ElBadry2018} appear to manifest only at higher masses ($M_{\ast}\sim 10^{7-8}\,M_{\sun}$). 
		
		\item Our dwarfs appear to under-predict the metallicities of observed LG satellites, with the discrepancy growing below $\mstar \ll 10^{7}\,M_{\sun}$ to just under 2 dex for UFDs. We argue that some of this may owe to pre-enrichment in the LG environment by the massive host, suggesting the observed LG dwarfs may not be universally representative. It may also reflect our neglect of any treatment of Pop-III stars (a single one of which could, in theory, produce more than enough metals to account for the discrepancy), or more generally the effects of more detailed progenitor mass and metallicity-dependent yields.
	\end{itemize}
	
	Overall, our study confirms many results previously found in lower-resolution simulations in the $\mstar \sim 10^3-10^6\mstar$ mass scale. Given that these simulations are beginning to resolve the evolution of individual supernovae, this suggests that the sub-grid approximations used in the lower-resolution simulations may be robust, at least insofar as their consequences for galaxy properties. The major uncertainties at the resolution level simulated here may no longer lie in sub-grid treatments of the collective effects of stellar feedback (which can be explicitly resolved) but rather in the fact that different individual stars have different evolutionary histories and feedback properties. One hopes that this may signal a turning point in simulations of dwarfs where we move from cosmological simulations with effective or sub-grid ISM physics models to those that directly model collapse and fragmentation of molecular clouds into individual stars, akin to previous studies of first stars or the rich studies of individual patches of the ISM (which actually often contain more mass than a UFD).

	\section*{Acknowledgments} 
	\vspace{-0.05cm}
	
	CW is supported by the Lee A. DuBridge Postdoctoral Scholarship in Astrophysics. Support for PFH and SGK was provided by an Alfred P. Sloan Research Fellowship, NSF Collaborative Research Grant \#1715847 and CAREER grant \#1455342, and NASA grants NNX15AT06G, JPL 1589742, 17-ATP17-0214. ABP acknowledges generous support from the George P. and Cynthia Woods Institute for Fundamental Physics and Astronomy at Texas A\&M University. SGK acknowledges additional support by NASA through Einstein Postdoctoral Fellowship grant number PF5-160136 awarded by the Chandra X-ray Center, which is operated by the Smithsonian Astrophysical Observatory for NASA under contract NAS8-03060. MBK acknowledges support from NSF grant AST-1517226 and CAREER grant AST-1752913 and from NASA grants NNX17AG29G and HST-AR-13888, HST-AR-13896, HST-AR-14282, HST-AR-14554, HST-AR-15006, HST-GO-12914, and HST-GO-14191 from the Space Telescope Science Institute, which is operated by AURA, Inc., under NASA contract NAS5-26555. CAFG was supported by NSF through grants AST-1517491, AST-1715216, and CAREER award AST-1652522, by NASA through grants NNX15AB22G and 17-ATP17-0067, and by a Cottrell Scholar Award from the Research Corporation for Science Advancement. Numerical calculations were run on the Caltech compute cluster ``Wheeler,'' allocations from XSEDE TG-AST130039 and PRAC NSF.1713353 supported by the NSF, and NASA HEC SMD-16-7592.
	\vspace{-0.5cm}

	\bibliography{dws_bib.bib}

\begin{thebibliography}{124}
\expandafter\ifx\csname natexlab\endcsname\relax\def\natexlab#1{#1}\fi

\bibitem[{{Amorisco}(2017)}]{Amorisco2017}
{Amorisco} N.~C., 2017, \apj, 844, 64

\bibitem[{{Applebaum} {et~al}\mbox{.}(2018){Applebaum}, {Brooks}, {Quinn}, \&
  {Christensen}}]{Applebaum2018}
{Applebaum} E., {Brooks} A.~M., {Quinn} T.~R., {Christensen} C.~R., 2018, ArXiv
  e-prints

\bibitem[{{Baushev} \& {Pilipenko}(2018)}]{Baushev2018}
{Baushev} A.~N., {Pilipenko} S.~V., 2018, ArXiv e-prints

\bibitem[{{Bechtol} {et~al}\mbox{.}(2015){Bechtol}, {Drlica-Wagner},
  {Balbinot}, {Pieres}, {Simon}, {Yanny}, {Santiago}, {Wechsler}, {Frieman},
  {Walker}, {Williams}, {Rozo}, {Rykoff}, {Queiroz}, {Luque},
  {Benoit-L{\'e}vy}, {Tucker}, {Sevilla}, {Gruendl}, {da Costa}, {Fausti Neto},
  {Maia}, {Abbott}, {Allam}, {Armstrong}, {Bauer}, {Bernstein}, {Bernstein},
  {Bertin}, {Brooks}, {Buckley-Geer}, {Burke}, {Carnero Rosell}, {Castander},
  {Covarrubias}, {D'Andrea}, {DePoy}, {Desai}, {Diehl}, {Eifler}, {Estrada},
  {Evrard}, {Fernandez}, {Finley}, {Flaugher}, {Gaztanaga}, {Gerdes},
  {Girardi}, {Gladders}, {Gruen}, {Gutierrez}, {Hao}, {Honscheid}, {Jain},
  {James}, {Kent}, {Kron}, {Kuehn}, {Kuropatkin}, {Lahav}, {Li}, {Lin},
  {Makler}, {March}, {Marshall}, {Martini}, {Merritt}, {Miller}, {Miquel},
  {Mohr}, {Neilsen}, {Nichol}, {Nord}, {Ogando}, {Peoples}, {Petravick},
  {Plazas}, {Romer}, {Roodman}, {Sako}, {Sanchez}, {Scarpine}, {Schubnell},
  {Smith}, {Soares-Santos}, {Sobreira}, {Suchyta}, {Swanson}, {Tarle},
  {Thaler}, {Thomas}, {Wester}, {Zuntz}, \& {DES Collaboration}}]{Bechtol2015}
{Bechtol} K. {et~al.}, 2015, \apj, 807, 50

\bibitem[{{Binney}(1978)}]{Binney1978}
{Binney} J., 1978, \mnras, 183, 501

\bibitem[{{Bovill} \& {Ricotti}(2011{\natexlab{a}})}]{Bovill2011a}
{Bovill} M.~S., {Ricotti} M., 2011{\natexlab{a}}, \apj, 741, 17

\bibitem[{{Bovill} \& {Ricotti}(2011{\natexlab{b}})}]{Bovill2011b}
{Bovill} M.~S., {Ricotti} M., 2011{\natexlab{b}}, \apj, 741, 18

\bibitem[{{Brook} {et~al}\mbox{.}(2014){Brook}, {Di Cintio}, {Knebe},
  {Gottl{\"o}ber}, {Hoffman}, {Yepes}, \& {Garrison-Kimmel}}]{Brook2014}
{Brook} C.~B., {Di Cintio} A., {Knebe} A., {Gottl{\"o}ber} S., {Hoffman} Y.,
  {Yepes} G., {Garrison-Kimmel} S., 2014, \apjl, 784, L14

\bibitem[{{Brown} {et~al}\mbox{.}(2014){Brown}, {Tumlinson}, {Geha}, {Simon},
  {Vargas}, {VandenBerg}, {Kirby}, {Kalirai}, {Avila}, {Gennaro}, {Ferguson},
  {Mu{\~n}oz}, {Guhathakurta}, \& {Renzini}}]{Brown2014}
{Brown} T.~M. {et~al.}, 2014, \apj, 796, 91

\bibitem[{{Bryan} \& {Norman}(1998)}]{Bryan1998}
{Bryan} G.~L., {Norman} M.~L., 1998, \apj, 495, 80

\bibitem[{{Buckley} \& {Peter}(2018)}]{buckley2018}
{Buckley} M.~R., {Peter} A.~H.~G., 2018, \physrep, 761, 1

\bibitem[{{Bullock}(2010)}]{Bullock2010b}
{Bullock} J.~S., 2010, ArXiv e-prints

\bibitem[{{Bullock} \& {Boylan-Kolchin}(2017)}]{Bullock2017}
{Bullock} J.~S., {Boylan-Kolchin} M., 2017, \araa, 55, 343

\bibitem[{{Bullock}, {Kravtsov} \& {Weinberg}(2000){Bullock}, {Kravtsov}, \&
  {Weinberg}}]{Bullock2000}
{Bullock} J.~S., {Kravtsov} A.~V., {Weinberg} D.~H., 2000, \apj, 539, 517

\bibitem[{{Bullock} {et~al}\mbox{.}(2010){Bullock}, {Stewart}, {Kaplinghat},
  {Tollerud}, \& {Wolf}}]{Bullock2010}
{Bullock} J.~S., {Stewart} K.~R., {Kaplinghat} M., {Tollerud} E.~J., {Wolf} J.,
  2010, \apj, 717, 1043

\bibitem[{{Caldwell} {et~al}\mbox{.}(2017){Caldwell}, {Strader}, {Sand},
  {Willman}, \& {Seth}}]{Caldwell2017}
{Caldwell} N., {Strader} J., {Sand} D.~J., {Willman} B., {Seth} A.~C., 2017,
  \mnras, 34, e039

\bibitem[{{Campbell} {et~al}\mbox{.}(2017){Campbell}, {Frenk}, {Jenkins},
  {Eke}, {Navarro}, {Sawala}, {Schaller}, {Fattahi}, {Oman}, \&
  {Theuns}}]{Campbell2017}
{Campbell} D.~J.~R. {et~al.}, 2017, \mnras, 469, 2335

\bibitem[{{Chan} {et~al}\mbox{.}(2015){Chan}, {Kere{\v s}}, {O{\~n}orbe},
  {Hopkins}, {Muratov}, {Faucher-Gigu{\`e}re}, \& {Quataert}}]{Chan2015}
{Chan} T.~K., {Kere{\v s}} D., {O{\~n}orbe} J., {Hopkins} P.~F., {Muratov}
  A.~L., {Faucher-Gigu{\`e}re} C.-A., {Quataert} E., 2015, \mnras, 454, 2981

\bibitem[{Chen(2015)}]{Chen2015}
Chen K.-J., 2015, Journal of Physics: Conference Series, 640, 012057

\bibitem[{{Cole} {et~al}\mbox{.}(2014){Cole}, {Weisz}, {Dolphin}, {Skillman},
  {McConnachie}, {Brooks}, \& {Leaman}}]{Cole2014}
{Cole} A.~A., {Weisz} D.~R., {Dolphin} A.~E., {Skillman} E.~D., {McConnachie}
  A.~W., {Brooks} A.~M., {Leaman} R., 2014, \apj, 795, 54

\bibitem[{{Collins} {et~al}\mbox{.}(2013){Collins}, {Chapman}, {Rich}, {Ibata},
  {Martin}, {Irwin}, {Bate}, {Lewis}, {Pe{\~n}arrubia}, {Arimoto}, {Casey},
  {Ferguson}, {Koch}, {McConnachie}, \& {Tanvir}}]{Collins2013}
{Collins} M.~L.~M. {et~al.}, 2013, \apj, 768, 172

\bibitem[{{de Blok}(2010)}]{deBlok2010}
{de Blok} W.~J.~G., 2010, Advances in Astronomy, 2010, 789293

\bibitem[{{de Blok} \& {Bosma}(2002)}]{deBlok2002}
{de Blok} W.~J.~G., {Bosma} A., 2002, \aap, 385, 816

\bibitem[{{Di Cintio} {et~al}\mbox{.}(2014{\natexlab{a}}){Di Cintio}, {Brook},
  {Macci{\`o}}, {Stinson}, {Knebe}, {Dutton}, \& {Wadsley}}]{DiCintio2014a}
{Di Cintio} A., {Brook} C.~B., {Macci{\`o}} A.~V., {Stinson} G.~S., {Knebe} A.,
  {Dutton} A.~A., {Wadsley} J., 2014{\natexlab{a}}, \mnras, 437, 415

\bibitem[{{Di Cintio} {et~al}\mbox{.}(2014{\natexlab{b}}){Di Cintio}, {Brook},
  {Macci{\`o}}, {Stinson}, {Knebe}, {Dutton}, \& {Wadsley}}]{diCintio2014}
{Di Cintio} A., {Brook} C.~B., {Macci{\`o}} A.~V., {Stinson} G.~S., {Knebe} A.,
  {Dutton} A.~A., {Wadsley} J., 2014{\natexlab{b}}, \mnras, 437, 415

\bibitem[{{Dijkstra} {et~al}\mbox{.}(2004){Dijkstra}, {Haiman}, {Rees}, \&
  {Weinberg}}]{Dijkstra2004}
{Dijkstra} M., {Haiman} Z., {Rees} M.~J., {Weinberg} D.~H., 2004, \apj, 601,
  666

\bibitem[{{Dooley} {et~al}\mbox{.}(2017){Dooley}, {Peter}, {Yang}, {Willman},
  {Griffen}, \& {Frebel}}]{Dooley2017}
{Dooley} G.~A., {Peter} A.~H.~G., {Yang} T., {Willman} B., {Griffen} B.~F.,
  {Frebel} A., 2017, \mnras, 471, 4894

\bibitem[{{Drlica-Wagner} {et~al}\mbox{.}(2015){Drlica-Wagner}, {Bechtol},
  {Rykoff}, {Luque}, {Queiroz}, {Mao}, {Wechsler}, {Simon}, {Santiago},
  {Yanny}, {Balbinot}, {Dodelson}, {Fausti Neto}, {James}, {Li}, {Maia},
  {Marshall}, {Pieres}, {Stringer}, {Walker}, {Abbott}, {Abdalla}, {Allam},
  {Benoit-L{\'e}vy}, {Bernstein}, {Bertin}, {Brooks}, {Buckley-Geer}, {Burke},
  {Carnero Rosell}, {Carrasco Kind}, {Carretero}, {Crocce}, {da Costa},
  {Desai}, {Diehl}, {Dietrich}, {Doel}, {Eifler}, {Evrard}, {Finley},
  {Flaugher}, {Fosalba}, {Frieman}, {Gaztanaga}, {Gerdes}, {Gruen}, {Gruendl},
  {Gutierrez}, {Honscheid}, {Kuehn}, {Kuropatkin}, {Lahav}, {Martini},
  {Miquel}, {Nord}, {Ogando}, {Plazas}, {Reil}, {Roodman}, {Sako}, {Sanchez},
  {Scarpine}, {Schubnell}, {Sevilla-Noarbe}, {Smith}, {Soares-Santos},
  {Sobreira}, {Suchyta}, {Swanson}, {Tarle}, {Tucker}, {Vikram}, {Wester},
  {Zhang}, {Zuntz}, \& {DES Collaboration}}]{DES2015}
{Drlica-Wagner} A. {et~al.}, 2015, \apj, 813, 109

\bibitem[{{Efstathiou}(1992)}]{Efstathiou1992}
{Efstathiou} G., 1992, \mnras, 256, 43P

\bibitem[{{Eisenstein} {et~al}\mbox{.}(2005){Eisenstein}, {Zehavi}, {Hogg},
  {Scoccimarro}, {Blanton}, {Nichol}, {Scranton}, {Seo}, {Tegmark}, {Zheng},
  {Anderson}, {Annis}, {Bahcall}, {Brinkmann}, {Burles}, {Castander},
  {Connolly}, {Csabai}, {Doi}, {Fukugita}, {Frieman}, {Glazebrook}, {Gunn},
  {Hendry}, {Hennessy}, {Ivezi{\'c}}, {Kent}, {Knapp}, {Lin}, {Loh}, {Lupton},
  {Margon}, {McKay}, {Meiksin}, {Munn}, {Pope}, {Richmond}, {Schlegel},
  {Schneider}, {Shimasaku}, {Stoughton}, {Strauss}, {SubbaRao}, {Szalay},
  {Szapudi}, {Tucker}, {Yanny}, \& {York}}]{Eisenstein2005}
{Eisenstein} D.~J. {et~al.}, 2005, \apj, 633, 560

\bibitem[{{El-Badry} {et~al}\mbox{.}(2018){El-Badry}, {Quataert}, {Wetzel},
  {Hopkins}, {Weisz}, {Chan}, {Fitts}, {Boylan-Kolchin}, {Kere{\v s}},
  {Faucher-Gigu{\`e}re}, \& {Garrison-Kimmel}}]{ElBadry2018}
{El-Badry} K. {et~al.}, 2018, \mnras, 473, 1930

\bibitem[{{El-Badry}, {Weisz} \& {Quataert}(2017){El-Badry}, {Weisz}, \&
  {Quataert}}]{ElBadry2017}
{El-Badry} K., {Weisz} D.~R., {Quataert} E., 2017, \mnras, 468, 319

\bibitem[{{El-Badry} {et~al}\mbox{.}(2017){El-Badry}, {Wetzel}, {Geha},
  {Quataert}, {Hopkins}, {Kere{\v s}}, {Chan}, \&
  {Faucher-Gigu{\`e}re}}]{ElBadry2017b}
{El-Badry} K., {Wetzel} A.~R., {Geha} M., {Quataert} E., {Hopkins} P.~F.,
  {Kere{\v s}} D., {Chan} T.~K., {Faucher-Gigu{\`e}re} C.-A., 2017, \apj, 835,
  193

\bibitem[{{Errani}, {Pe{\~n}arrubia} \& {Walker}(2018){Errani},
  {Pe{\~n}arrubia}, \& {Walker}}]{Errani2018}
{Errani} R., {Pe{\~n}arrubia} J., {Walker} M.~G., 2018, \mnras, 481, 5073

\bibitem[{{Escala} {et~al}\mbox{.}(2018){Escala}, {Wetzel}, {Kirby}, {Hopkins},
  {Ma}, {Wheeler}, {Kere{\v s}}, {Faucher-Gigu{\`e}re}, \&
  {Quataert}}]{Escala2018}
{Escala} I. {et~al.}, 2018, \mnras, 474, 2194

\bibitem[{{Faucher-Gigu{\`e}re}(2018)}]{Faucher-Giguere2018}
{Faucher-Gigu{\`e}re} C.-A., 2018, \mnras, 473, 3717

\bibitem[{{Faucher-Gigu{\`e}re} {et~al}\mbox{.}(2009){Faucher-Gigu{\`e}re},
  {Lidz}, {Zaldarriaga}, \& {Hernquist}}]{Faucher-Guigere2009}
{Faucher-Gigu{\`e}re} C.-A., {Lidz} A., {Zaldarriaga} M., {Hernquist} L., 2009,
  \apj, 703, 1416

\bibitem[{{Fillingham} {et~al}\mbox{.}(2018){Fillingham}, {Cooper},
  {Boylan-Kolchin}, {Bullock}, {Garrison-Kimmel}, \&
  {Wheeler}}]{Fillingham2018}
{Fillingham} S.~P., {Cooper} M.~C., {Boylan-Kolchin} M., {Bullock} J.~S.,
  {Garrison-Kimmel} S., {Wheeler} C., 2018, \mnras, 477, 4491

\bibitem[{{Fitts} {et~al}\mbox{.}(2017){Fitts}, {Boylan-Kolchin}, {Elbert},
  {Bullock}, {Hopkins}, {O{\~n}orbe}, {Wetzel}, {Wheeler},
  {Faucher-Gigu{\`e}re}, {Kere{\v s}}, {Skillman}, \& {Weisz}}]{Fitts2017}
{Fitts} A. {et~al.}, 2017, \mnras, 471, 3547

\bibitem[{{Flores} \& {Primack}(1994)}]{Flores1994}
{Flores} R.~A., {Primack} J.~R., 1994, \apjl, 427, L1

\bibitem[{{Fran{\c c}ois} {et~al}\mbox{.}(2004){Fran{\c c}ois}, {Matteucci},
  {Cayrel}, {Spite}, {Spite}, \& {Chiappini}}]{Francois2004}
{Fran{\c c}ois} P., {Matteucci} F., {Cayrel} R., {Spite} M., {Spite} F.,
  {Chiappini} C., 2004, \aap, 421, 613

\bibitem[{{Garrison-Kimmel} {et~al}\mbox{.}(2014){Garrison-Kimmel},
  {Boylan-Kolchin}, {Bullock}, \& {Lee}}]{Garrison-Kimmel2014a}
{Garrison-Kimmel} S., {Boylan-Kolchin} M., {Bullock} J.~S., {Lee} K., 2014,
  \mnras, 438, 2578

\bibitem[{{Garrison-Kimmel} {et~al}\mbox{.}(2018){Garrison-Kimmel}, {Hopkins},
  {Wetzel}, {Bullock}, {Boylan-Kolchin}, {Keres}, {Faucher-Giguere},
  {El-Badry}, {Lamberts}, {Quataert}, \& {Sanderson}}]{Garrison-Kimmel2018}
{Garrison-Kimmel} S. {et~al.}, 2018, ArXiv e-prints

\bibitem[{{Garrison-Kimmel} {et~al}\mbox{.}(2013){Garrison-Kimmel}, {Rocha},
  {Boylan-Kolchin}, {Bullock}, \& {Lally}}]{Garrison-Kimmel2013a}
{Garrison-Kimmel} S., {Rocha} M., {Boylan-Kolchin} M., {Bullock} J.~S., {Lally}
  J., 2013, \mnras, 433, 3539

\bibitem[{{Garrison-Kimmel} {et~al}\mbox{.}(2017){Garrison-Kimmel}, {Wetzel},
  {Bullock}, {Hopkins}, {Boylan-Kolchin}, {Faucher-Gigu{\`e}re}, {Kere{\v s}},
  {Quataert}, {Sanderson}, {Graus}, \& {Kelley}}]{Garrison-Kimmel2017}
{Garrison-Kimmel} S. {et~al.}, 2017, \mnras, 471, 1709

\bibitem[{{Geha} {et~al}\mbox{.}(2012){Geha}, {Blanton}, {Yan}, \&
  {Tinker}}]{Geha:2012kx}
{Geha} M., {Blanton} M.~R., {Yan} R., {Tinker} J.~L., 2012, \apj, 757, 85

\bibitem[{{Goerdt} {et~al}\mbox{.}(2006){Goerdt}, {Moore}, {Read}, {Stadel}, \&
  {Zemp}}]{Goerdt2006}
{Goerdt} T., {Moore} B., {Read} J.~I., {Stadel} J., {Zemp} M., 2006, \mnras,
  368, 1073

\bibitem[{{Gonz{\'a}lez-Samaniego}
  {et~al}\mbox{.}(2017){Gonz{\'a}lez-Samaniego}, {Bullock}, {Boylan-Kolchin},
  {Fitts}, {Elbert}, {Hopkins}, {Kere{\v s}}, \&
  {Faucher-Gigu{\`e}re}}]{Gonzalez2017}
{Gonz{\'a}lez-Samaniego} A., {Bullock} J.~S., {Boylan-Kolchin} M., {Fitts} A.,
  {Elbert} O.~D., {Hopkins} P.~F., {Kere{\v s}} D., {Faucher-Gigu{\`e}re}
  C.-A., 2017, \mnras, 472, 4786

\bibitem[{{Governato} {et~al}\mbox{.}(2012){Governato}, {Zolotov}, {Pontzen},
  {Christensen}, {Oh}, {Brooks}, {Quinn}, {Shen}, \&
  {Wadsley}}]{Governato2012G}
{Governato} F. {et~al.}, 2012, \mnras, 422, 1231

\bibitem[{{Graus} {et~al}\mbox{.}(2018){Graus}, {Bullock}, {Kelley},
  {Boylan-Kolchin}, {Garrison-Kimmel}, \& {Qi}}]{Graus2018}
{Graus} A.~S., {Bullock} J.~S., {Kelley} T., {Boylan-Kolchin} M.,
  {Garrison-Kimmel} S., {Qi} Y., 2018, ArXiv e-prints

\bibitem[{{Grudi{\'c}} \& {Hopkins}(2018)}]{Grudic2018}
{Grudi{\'c}} M.~Y., {Hopkins} P.~F., 2018, ArXiv e-prints

\bibitem[{{Gunn} \& {Gott}(1972)}]{Gunn1972}
{Gunn} J.~E., {Gott}, III J.~R., 1972, \apj, 176, 1

\bibitem[{{Hoeft} {et~al}\mbox{.}(2006){Hoeft}, {Yepes}, {Gottl{\"o}ber}, \&
  {Springel}}]{Hoeft2006}
{Hoeft} M., {Yepes} G., {Gottl{\"o}ber} S., {Springel} V., 2006, \mnras, 371,
  401

\bibitem[{{Homma} {et~al}\mbox{.}(2016){Homma}, {Chiba}, {Okamoto}, {Komiyama},
  {Tanaka}, {Tanaka}, {Ishigaki}, {Akiyama}, {Arimoto}, {Garmilla}, {Lupton},
  {Strauss}, {Furusawa}, {Miyazaki}, {Murayama}, {Nishizawa}, {Takada},
  {Usuda}, \& {Wang}}]{Homma2016}
{Homma} D. {et~al.}, 2016, \apj, 832, 21

\bibitem[{{Homma} {et~al}\mbox{.}(2018){Homma}, {Chiba}, {Okamoto}, {Komiyama},
  {Tanaka}, {Tanaka}, {Ishigaki}, {Hayashi}, {Arimoto}, {Garmilla}, {Lupton},
  {Strauss}, {Miyazaki}, {Wang}, \& {Murayama}}]{Homma2018}
{Homma} D. {et~al.}, 2018, \pasj, 70, S18

\bibitem[{{Hopkins}(2014)}]{Hopkins2014b}
{Hopkins} P.~F., 2014, ArXiv e-prints

\bibitem[{{Hopkins} {et~al}\mbox{.}(2014){Hopkins}, {Kere{\v s}}, {O{\~n}orbe},
  {Faucher-Gigu{\`e}re}, {Quataert}, {Murray}, \& {Bullock}}]{Hopkins2014a}
{Hopkins} P.~F., {Kere{\v s}} D., {O{\~n}orbe} J., {Faucher-Gigu{\`e}re} C.-A.,
  {Quataert} E., {Murray} N., {Bullock} J.~S., 2014, \mnras, 445, 581

\bibitem[{{Hopkins} {et~al}\mbox{.}(2018{\natexlab{a}}){Hopkins}, {Wetzel},
  {Kere{\v s}}, {Faucher-Gigu{\`e}re}, {Quataert}, {Boylan-Kolchin}, {Murray},
  {Hayward}, \& {El-Badry}}]{Hopkins2018b}
{Hopkins} P.~F. {et~al.}, 2018{\natexlab{a}}, \mnras, 477, 1578

\bibitem[{{Hopkins} {et~al}\mbox{.}(2018{\natexlab{b}}){Hopkins}, {Wetzel},
  {Kere{\v s}}, {Faucher-Gigu{\`e}re}, {Quataert}, {Boylan-Kolchin}, {Murray},
  {Hayward}, {Garrison-Kimmel}, {Hummels}, {Feldmann}, {Torrey}, {Ma},
  {Angl{\'e}s-Alc{\'a}zar}, {Su}, {Orr}, {Schmitz}, {Escala}, {Sanderson},
  {Grudi{\'c}}, {Hafen}, {Kim}, {Fitts}, {Bullock}, {Wheeler}, {Chan},
  {Elbert}, \& {Narayanan}}]{Hopkins2017}
{Hopkins} P.~F. {et~al.}, 2018{\natexlab{b}}, \mnras, 480, 800

\bibitem[{{Hu}(2018)}]{Hu2018}
{Hu} C.-Y., 2018, ArXiv e-prints

\bibitem[{{Jaacks} {et~al}\mbox{.}(2018){Jaacks}, {Thompson}, {Finkelstein}, \&
  {Bromm}}]{Jaacks2018}
{Jaacks} J., {Thompson} R., {Finkelstein} S.~L., {Bromm} V., 2018, \mnras, 475,
  4396

\bibitem[{{Karachentsev} {et~al}\mbox{.}(2015){Karachentsev}, {Makarova},
  {Makarov}, {Tully}, \& {Rizzi}}]{Karachentsev2015}
{Karachentsev} I.~D., {Makarova} L.~N., {Makarov} D.~I., {Tully} R.~B., {Rizzi}
  L., 2015, \mnras, 447, L85

\bibitem[{{Karachentsev} {et~al}\mbox{.}(2001){Karachentsev}, {Sharina},
  {Dolphin}, {Geisler}, {Grebel}, {Guhathakurta}, {Hodge}, {Karachentseva},
  {Sarajedini}, \& {Seitzer}}]{Karachentsev2001}
{Karachentsev} I.~D. {et~al.}, 2001, \aap, 379, 407

\bibitem[{{Karachentsev} {et~al}\mbox{.}(2014){Karachentsev}, {Tully}, {Wu},
  {Shaya}, \& {Dolphin}}]{Karachentsev2014}
{Karachentsev} I.~D., {Tully} R.~B., {Wu} P.-F., {Shaya} E.~J., {Dolphin}
  A.~E., 2014, \apj, 782, 4

\bibitem[{{Kaufmann}, {Wheeler} \& {Bullock}(2007){Kaufmann}, {Wheeler}, \&
  {Bullock}}]{Kaufmann2007}
{Kaufmann} T., {Wheeler} C., {Bullock} J.~S., 2007, \mnras, 382, 1187

\bibitem[{{Keller} {et~al}\mbox{.}(2019){Keller}, {Wadsley}, {Wang}, \&
  {Kruijssen}}]{Keller2019}
{Keller} B.~W., {Wadsley} J.~W., {Wang} L., {Kruijssen} J.~M.~D., 2019, \mnras,
  482, 2244

\bibitem[{{Kim} \& {Ostriker}(2015)}]{Kim2015c}
{Kim} C.-G., {Ostriker} E.~C., 2015, \apj, 802, 99

\bibitem[{{Kirby} {et~al}\mbox{.}(2013{\natexlab{a}}){Kirby}, {Boylan-Kolchin},
  {Cohen}, {Geha}, {Bullock}, \& {Kaplinghat}}]{Kirby2013}
{Kirby} E.~N., {Boylan-Kolchin} M., {Cohen} J.~G., {Geha} M., {Bullock} J.~S.,
  {Kaplinghat} M., 2013{\natexlab{a}}, \apj, 770, 16

\bibitem[{{Kirby} {et~al}\mbox{.}(2014){Kirby}, {Bullock}, {Boylan-Kolchin},
  {Kaplinghat}, \& {Cohen}}]{Kirby2014}
{Kirby} E.~N., {Bullock} J.~S., {Boylan-Kolchin} M., {Kaplinghat} M., {Cohen}
  J.~G., 2014, \mnras, 439, 1015

\bibitem[{{Kirby} {et~al}\mbox{.}(2013{\natexlab{b}}){Kirby}, {Cohen},
  {Guhathakurta}, {Cheng}, {Bullock}, \& {Gallazzi}}]{Kirby2013A}
{Kirby} E.~N., {Cohen} J.~G., {Guhathakurta} P., {Cheng} L., {Bullock} J.~S.,
  {Gallazzi} A., 2013{\natexlab{b}}, \apj, 779, 102

\bibitem[{{Kirby}, {Martin} \& {Finlator}(2011){Kirby}, {Martin}, \&
  {Finlator}}]{Kirby2011}
{Kirby} E.~N., {Martin} C.~L., {Finlator} K., 2011, \apjl, 742, L25

\bibitem[{{Klypin} {et~al}\mbox{.}(1999){Klypin}, {Kravtsov}, {Valenzuela}, \&
  {Prada}}]{Klypin1999}
{Klypin} A., {Kravtsov} A.~V., {Valenzuela} O., {Prada} F., 1999, \apj, 522, 82

\bibitem[{{Knollmann} \& {Knebe}(2009)}]{Knollmann2009}
{Knollmann} S.~R., {Knebe} A., 2009, \apjs, 182, 608

\bibitem[{{Komatsu} {et~al}\mbox{.}(2011){Komatsu}, {Smith}, {Dunkley},
  {Bennett}, {Gold}, {Hinshaw}, {Jarosik}, {Larson}, {Nolta}, {Page},
  {Spergel}, {Halpern}, {Hill}, {Kogut}, {Limon}, {Meyer}, {Odegard}, {Tucker},
  {Weiland}, {Wollack}, \& {Wright}}]{Komatsu2011}
{Komatsu} E. {et~al.}, 2011, \apjs, 192, 18

\bibitem[{{Kozyreva}, {Yoon} \& {Langer}(2014){Kozyreva}, {Yoon}, \&
  {Langer}}]{Kozyreva2014}
{Kozyreva} A., {Yoon} S.-C., {Langer} N., 2014, \aap, 566, A146

\bibitem[{{Kravtsov}(2013)}]{Kravtsov2013}
{Kravtsov} A.~V., 2013, \apjl, 764, L31

\bibitem[{{Kroupa}(2002)}]{Kroupa2002}
{Kroupa} P., 2002, Science, 295, 82

\bibitem[{{Lapi}, {Cavaliere} \& {Menci}(2005){Lapi}, {Cavaliere}, \&
  {Menci}}]{Lapi2005}
{Lapi} A., {Cavaliere} A., {Menci} N., 2005, \apj, 619, 60

\bibitem[{{Leitherer} {et~al}\mbox{.}(1999){Leitherer}, {Schaerer}, {Goldader},
  {Delgado}, {Robert}, {Kune}, {de Mello}, {Devost}, \&
  {Heckman}}]{Leitherer1999}
{Leitherer} C. {et~al.}, 1999, \apjs, 123, 3

\bibitem[{{Ma} {et~al}\mbox{.}(2017{\natexlab{a}}){Ma}, {Hopkins}, {Feldmann},
  {Torrey}, {Faucher-Gigu{\`e}re}, \& {Kere{\v s}}}]{Ma2017a}
{Ma} X., {Hopkins} P.~F., {Feldmann} R., {Torrey} P., {Faucher-Gigu{\`e}re}
  C.-A., {Kere{\v s}} D., 2017{\natexlab{a}}, \mnras, 466, 4780

\bibitem[{{Ma} {et~al}\mbox{.}(2017{\natexlab{b}}){Ma}, {Hopkins}, {Wetzel},
  {Kirby}, {Angl{\'e}s-Alc{\'a}zar}, {Faucher-Gigu{\`e}re}, {Kere{\v s}}, \&
  {Quataert}}]{Ma2017b}
{Ma} X., {Hopkins} P.~F., {Wetzel} A.~R., {Kirby} E.~N.,
  {Angl{\'e}s-Alc{\'a}zar} D., {Faucher-Gigu{\`e}re} C.-A., {Kere{\v s}} D.,
  {Quataert} E., 2017{\natexlab{b}}, \mnras, 467, 2430

\bibitem[{{Ma} {et~al}\mbox{.}(2015){Ma}, {Kasen}, {Hopkins},
  {Faucher-Gigu{\`e}re}, {Quataert}, {Kere{\v s}}, \& {Murray}}]{Ma2015}
{Ma} X., {Kasen} D., {Hopkins} P.~F., {Faucher-Gigu{\`e}re} C.-A., {Quataert}
  E., {Kere{\v s}} D., {Murray} N., 2015, \mnras, 453, 960

\bibitem[{{Makarov} {et~al}\mbox{.}(2012){Makarov}, {Makarova}, {Sharina},
  {Uklein}, {Tikhonov}, {Guhathakurta}, {Kirby}, \& {Terekhova}}]{Makarov2012}
{Makarov} D., {Makarova} L., {Sharina} M., {Uklein} R., {Tikhonov} A.,
  {Guhathakurta} P., {Kirby} E., {Terekhova} N., 2012, \mnras, 425, 709

\bibitem[{{Makarova} {et~al}\mbox{.}(2017){Makarova}, {Makarov},
  {Karachentsev}, {Tully}, \& {Rizzi}}]{Makarov2017}
{Makarova} L.~N., {Makarov} D.~I., {Karachentsev} I.~D., {Tully} R.~B., {Rizzi}
  L., 2017, \mnras, 464, 2281

\bibitem[{{Martin} {et~al}\mbox{.}(2015){Martin}, {Nidever}, {Besla}, {Olsen},
  {Walker}, {Vivas}, {Gruendl}, {Kaleida}, {Mu{\~n}oz}, {Blum}, {Saha}, {Conn},
  {Bell}, {Chu}, {Cioni}, {de Boer}, {Gallart}, {Jin}, {Kunder}, {Majewski},
  {Martinez-Delgado}, {Monachesi}, {Monelli}, {Monteagudo}, {No{\"e}l},
  {Olszewski}, {Stringfellow}, {van der Marel}, \& {Zaritsky}}]{Martin2015}
{Martin} N.~F. {et~al.}, 2015, \apjl, 804, L5

\bibitem[{{Martizzi}, {Faucher-Gigu{\`e}re} \& {Quataert}(2015){Martizzi},
  {Faucher-Gigu{\`e}re}, \& {Quataert}}]{Martizzi2015}
{Martizzi} D., {Faucher-Gigu{\`e}re} C.-A., {Quataert} E., 2015, \mnras, 450,
  504

\bibitem[{{Mashchenko}, {Couchman} \& {Wadsley}(2006){Mashchenko}, {Couchman},
  \& {Wadsley}}]{Mashchenko2006}
{Mashchenko} S., {Couchman} H.~M.~P., {Wadsley} J., 2006, \nat, 442, 539

\bibitem[{{McConnachie}(2012)}]{McConnachie2012}
{McConnachie} A.~W., 2012, \aj, 144, 4

\bibitem[{{Mo}, {Mao} \& {White}(1998){Mo}, {Mao}, \& {White}}]{MMW1998}
{Mo} H.~J., {Mao} S., {White} S.~D.~M., 1998, \mnras, 295, 319

\bibitem[{{Moore}(1994)}]{Moore1994}
{Moore} B., 1994, \nat, 370, 629

\bibitem[{{Moore} {et~al}\mbox{.}(1999){Moore}, {Ghigna}, {Governato}, {Lake},
  {Quinn}, {Stadel}, \& {Tozzi}}]{Moore1999}
{Moore} B., {Ghigna} S., {Governato} F., {Lake} G., {Quinn} T., {Stadel} J.,
  {Tozzi} P., 1999, \apjl, 524, L19

\bibitem[{{Munshi} {et~al}\mbox{.}(2017){Munshi}, {Brooks}, {Applebaum},
  {Weisz}, {Governato}, \& {Quinn}}]{Munshi2017}
{Munshi} F., {Brooks} A.~M., {Applebaum} E., {Weisz} D.~R., {Governato} F.,
  {Quinn} T.~R., 2017, ArXiv e-prints

\bibitem[{{Munshi} {et~al}\mbox{.}(2013){Munshi}, {Governato}, {Brooks},
  {Christensen}, {Shen}, {Loebman}, {Moster}, {Quinn}, \&
  {Wadsley}}]{Munshi2013}
{Munshi} F. {et~al.}, 2013, \apj, 766, 56

\bibitem[{{Muratov} {et~al}\mbox{.}(2015){Muratov}, {Kere{\v s}},
  {Faucher-Gigu{\`e}re}, {Hopkins}, {Quataert}, \& {Murray}}]{Muratov2015}
{Muratov} A.~L., {Kere{\v s}} D., {Faucher-Gigu{\`e}re} C.-A., {Hopkins} P.~F.,
  {Quataert} E., {Murray} N., 2015, \mnras, 454, 2691

\bibitem[{{Navarro}, {Frenk} \& {White}(1997){Navarro}, {Frenk}, \&
  {White}}]{NFW1997}
{Navarro} J.~F., {Frenk} C.~S., {White} S.~D.~M., 1997, \apj, 490, 493

\bibitem[{{O{\~n}orbe} {et~al}\mbox{.}(2015){O{\~n}orbe}, {Boylan-Kolchin},
  {Bullock}, {Hopkins}, {Kere{\v s}}, {Faucher-Gigu{\`e}re}, {Quataert}, \&
  {Murray}}]{Onorbe2015}
{O{\~n}orbe} J., {Boylan-Kolchin} M., {Bullock} J.~S., {Hopkins} P.~F.,
  {Kere{\v s}} D., {Faucher-Gigu{\`e}re} C.-A., {Quataert} E., {Murray} N.,
  2015, \mnras, 454, 2092

\bibitem[{{Oh} {et~al}\mbox{.}(2008){Oh}, {de Blok}, {Walter}, {Brinks}, \&
  {Kennicutt}}]{Oh2008}
{Oh} S.-H., {de Blok} W.~J.~G., {Walter} F., {Brinks} E., {Kennicutt}, Jr.
  R.~C., 2008, \aj, 136, 2761

\bibitem[{{Planck Collaboration} {et~al}\mbox{.}(2014){Planck Collaboration},
  {Ade}, {Aghanim}, {Armitage-Caplan}, {Arnaud}, {Ashdown}, {Atrio-Barandela},
  {Aumont}, {Baccigalupi}, {Banday}, \& et~al.}]{Planck2014}
{Planck Collaboration} {et~al.}, 2014, \aap, 571, A16

\bibitem[{{Pontzen} \& {Governato}(2012)}]{Pontzen2012}
{Pontzen} A., {Governato} F., 2012, \mnras, 421, 3464

\bibitem[{{Power} {et~al}\mbox{.}(2003){Power}, {Navarro}, {Jenkins}, {Frenk},
  {White}, {Springel}, {Stadel}, \& {Quinn}}]{Power2003}
{Power} C., {Navarro} J.~F., {Jenkins} A., {Frenk} C.~S., {White} S.~D.~M.,
  {Springel} V., {Stadel} J., {Quinn} T., 2003, \mnras, 338, 14

\bibitem[{{Read}, {Agertz} \& {Collins}(2016){Read}, {Agertz}, \&
  {Collins}}]{Read2016}
{Read} J.~I., {Agertz} O., {Collins} M.~L.~M., 2016, \mnras, 459, 2573

\bibitem[{{Reid} {et~al}\mbox{.}(2010){Reid}, {Percival}, {Eisenstein},
  {Verde}, {Spergel}, {Skibba}, {Bahcall}, {Budavari}, {Frieman}, {Fukugita},
  {Gott}, {Gunn}, {Ivezi{\'c}}, {Knapp}, {Kron}, {Lupton}, {McKay}, {Meiksin},
  {Nichol}, {Pope}, {Schlegel}, {Schneider}, {Stoughton}, {Strauss}, {Szalay},
  {Tegmark}, {Vogeley}, {Weinberg}, {York}, \& {Zehavi}}]{Reid2010}
{Reid} B.~A. {et~al.}, 2010, \mnras, 404, 60

\bibitem[{{Ricotti} \& {Gnedin}(2005)}]{Ricotti2005}
{Ricotti} M., {Gnedin} N.~Y., 2005, \apj, 629, 259

\bibitem[{{Rodriguez Wimberly} {et~al}\mbox{.}(2018){Rodriguez Wimberly},
  {Cooper}, {Fillingham}, {Boylan-Kolchin}, {Bullock}, \&
  {Garrison-Kimmel}}]{Rodriguez-Wimberly2018}
{Rodriguez Wimberly} M.~K., {Cooper} M.~C., {Fillingham} S.~P.,
  {Boylan-Kolchin} M., {Bullock} J.~S., {Garrison-Kimmel} S., 2018, ArXiv
  e-prints

\bibitem[{{Salucci} \& {Burkert}(2000)}]{Salucci2000}
{Salucci} P., {Burkert} A., 2000, \apjl, 537, L9

\bibitem[{{Sawala} {et~al}\mbox{.}(2014){Sawala}, {Frenk}, {Fattahi},
  {Navarro}, {Theuns}, {Bower}, {Crain}, {Furlong}, {Jenkins}, {Schaller}, \&
  {Schaye}}]{Sawala2014}
{Sawala} T. {et~al.}, 2014, ArXiv e-prints

\bibitem[{{Sparre} {et~al}\mbox{.}(2017){Sparre}, {Hayward}, {Feldmann},
  {Faucher-Gigu{\`e}re}, {Muratov}, {Kere{\v s}}, \& {Hopkins}}]{Sparre2017}
{Sparre} M., {Hayward} C.~C., {Feldmann} R., {Faucher-Gigu{\`e}re} C.-A.,
  {Muratov} A.~L., {Kere{\v s}} D., {Hopkins} P.~F., 2017, \mnras, 466, 88

\bibitem[{{Su} {et~al}\mbox{.}(2018{\natexlab{a}}){Su}, {Hayward}, {Hopkins},
  {Quataert}, {Faucher-Gigu{\`e}re}, \& {Kere{\v s}}}]{Su2018a}
{Su} K.-Y., {Hayward} C.~C., {Hopkins} P.~F., {Quataert} E.,
  {Faucher-Gigu{\`e}re} C.-A., {Kere{\v s}} D., 2018{\natexlab{a}}, \mnras,
  473, L111

\bibitem[{{Su} {et~al}\mbox{.}(2018{\natexlab{b}}){Su}, {Hopkins}, {Hayward},
  {Ma}, {Boylan-Kolchin}, {Kasen}, {Kere{\v s}}, {Faucher-Gigu{\`e}re}, {Orr},
  \& {Wheeler}}]{Su2018}
{Su} K.-Y. {et~al.}, 2018{\natexlab{b}}, \mnras, 480, 1666

\bibitem[{{Tollet} {et~al}\mbox{.}(2016){Tollet}, {Macci{\`o}}, {Dutton},
  {Stinson}, {Wang}, {Penzo}, {Gutcke}, {Buck}, {Kang}, {Brook}, {Di Cintio},
  {Keller}, \& {Wadsley}}]{tollet2016}
{Tollet} E. {et~al.}, 2016, \mnras, 456, 3542

\bibitem[{{Torrealba} {et~al}\mbox{.}(2018){Torrealba}, {Belokurov}, {Koposov},
  {Li}, {Walker}, {Sanders}, {Geringer-Sameth}, {Zucker}, {Kuehn}, {Evans}, \&
  {Dehnen}}]{Torrealba2018}
{Torrealba} G. {et~al.}, 2018, ArXiv e-prints

\bibitem[{{Tulin} \& {Yu}(2018)}]{tulin2018}
{Tulin} S., {Yu} H.-B., 2018, \physrep, 730, 1

\bibitem[{{van den Bosch} \& {Dalcanton}(2000)}]{vandenBosch2000}
{van den Bosch} F.~C., {Dalcanton} J.~J., 2000, \apj, 534, 146

\bibitem[{{van Dokkum} {et~al}\mbox{.}(2015){van Dokkum}, {Abraham}, {Merritt},
  {Zhang}, {Geha}, \& {Conroy}}]{vanDokkum2015}
{van Dokkum} P.~G., {Abraham} R., {Merritt} A., {Zhang} J., {Geha} M., {Conroy}
  C., 2015, \apjl, 798, L45

\bibitem[{{Vargas}, {Geha} \& {Tollerud}(2014){Vargas}, {Geha}, \&
  {Tollerud}}]{Vargas2014}
{Vargas} L.~C., {Geha} M.~C., {Tollerud} E.~J., 2014, \apj, 790, 73

\bibitem[{{Viel} {et~al}\mbox{.}(2008){Viel}, {Becker}, {Bolton}, {Haehnelt},
  {Rauch}, \& {Sargent}}]{Viel2008}
{Viel} M., {Becker} G.~D., {Bolton} J.~S., {Haehnelt} M.~G., {Rauch} M.,
  {Sargent} W.~L.~W., 2008, Physical Review Letters, 100, 041304

\bibitem[{{Walch} \& {Naab}(2015)}]{Walch2015}
{Walch} S., {Naab} T., 2015, \mnras, 451, 2757

\bibitem[{{Weinmann} {et~al}\mbox{.}(2007){Weinmann}, {Macci{\`o}}, {Iliev},
  {Mellema}, \& {Moore}}]{Weinmann2007}
{Weinmann} S.~M., {Macci{\`o}} A.~V., {Iliev} I.~T., {Mellema} G., {Moore} B.,
  2007, \mnras, 381, 367

\bibitem[{{Wetzel} {et~al}\mbox{.}(2016){Wetzel}, {Hopkins}, {Kim},
  {Faucher-Gigu{\`e}re}, {Kere{\v s}}, \& {Quataert}}]{Wetzel2016}
{Wetzel} A.~R., {Hopkins} P.~F., {Kim} J.-h., {Faucher-Gigu{\`e}re} C.-A.,
  {Kere{\v s}} D., {Quataert} E., 2016, \apjl, 827, L23

\bibitem[{{Wheeler} {et~al}\mbox{.}(2015){Wheeler}, {O{\~n}orbe}, {Bullock},
  {Boylan-Kolchin}, {Elbert}, {Garrison-Kimmel}, {Hopkins}, \& {Kere{\v
  s}}}]{Wheeler2015}
{Wheeler} C., {O{\~n}orbe} J., {Bullock} J.~S., {Boylan-Kolchin} M., {Elbert}
  O.~D., {Garrison-Kimmel} S., {Hopkins} P.~F., {Kere{\v s}} D., 2015, \mnras,
  453, 1305

\bibitem[{{Wheeler} {et~al}\mbox{.}(2017){Wheeler}, {Pace}, {Bullock},
  {Boylan-Kolchin}, {O{\~n}orbe}, {Elbert}, {Fitts}, {Hopkins}, \& {Kere{\v
  s}}}]{Wheeler2017}
{Wheeler} C. {et~al.}, 2017, \mnras, 465, 2420

\bibitem[{{Wolf} {et~al}\mbox{.}(2010){Wolf}, {Martinez}, {Bullock},
  {Kaplinghat}, {Geha}, {Mu{\~n}oz}, {Simon}, \& {Avedo}}]{Wolf2010}
{Wolf} J., {Martinez} G.~D., {Bullock} J.~S., {Kaplinghat} M., {Geha} M.,
  {Mu{\~n}oz} R.~R., {Simon} J.~D., {Avedo} F.~F., 2010, \mnras, 406, 1220

\bibitem[{{Woosley} \& {Weaver}(1995)}]{Woosley1995}
{Woosley} S.~E., {Weaver} T.~A., 1995, \apjs, 101, 181

\bibitem[{{York} {et~al}\mbox{.}(2000){York}, {Adelman}, {Anderson},
  {Anderson}, {Annis}, {Bahcall}, {Bakken}, {Barkhouser}, {Bastian}, {Berman},
  {Boroski}, {Bracker}, {Briegel}, {Briggs}, {Brinkmann}, {Brunner}, {Burles},
  {Carey}, {Carr}, {Castander}, {Chen}, {Colestock}, {Connolly}, {Crocker},
  {Csabai}, {Czarapata}, {Davis}, {Doi}, {Dombeck}, {Eisenstein}, {Ellman},
  {Elms}, {Evans}, {Fan}, {Federwitz}, {Fiscelli}, {Friedman}, {Frieman},
  {Fukugita}, {Gillespie}, {Gunn}, {Gurbani}, {de Haas}, {Haldeman}, {Harris},
  {Hayes}, {Heckman}, {Hennessy}, {Hindsley}, {Holm}, {Holmgren}, {Huang},
  {Hull}, {Husby}, {Ichikawa}, {Ichikawa}, {Ivezi{\'c}}, {Kent}, {Kim},
  {Kinney}, {Klaene}, {Kleinman}, {Kleinman}, {Knapp}, {Korienek}, {Kron},
  {Kunszt}, {Lamb}, {Lee}, {Leger}, {Limmongkol}, {Lindenmeyer}, {Long},
  {Loomis}, {Loveday}, {Lucinio}, {Lupton}, {MacKinnon}, {Mannery}, {Mantsch},
  {Margon}, {McGehee}, {McKay}, {Meiksin}, {Merelli}, {Monet}, {Munn},
  {Narayanan}, {Nash}, {Neilsen}, {Neswold}, {Newberg}, {Nichol}, {Nicinski},
  {Nonino}, {Okada}, {Okamura}, {Ostriker}, {Owen}, {Pauls}, {Peoples},
  {Peterson}, {Petravick}, {Pier}, {Pope}, {Pordes}, {Prosapio},
  {Rechenmacher}, {Quinn}, {Richards}, {Richmond}, {Rivetta}, {Rockosi},
  {Ruthmansdorfer}, {Sandford}, {Schlegel}, {Schneider}, {Sekiguchi}, {Sergey},
  {Shimasaku}, {Siegmund}, {Smee}, {Smith}, {Snedden}, {Stone}, {Stoughton},
  {Strauss}, {Stubbs}, {SubbaRao}, {Szalay}, {Szapudi}, {Szokoly}, {Thakar},
  {Tremonti}, {Tucker}, {Uomoto}, {Vanden Berk}, {Vogeley}, {Waddell}, {Wang},
  {Watanabe}, {Weinberg}, {Yanny}, {Yasuda}, \& {SDSS Collaboration}}]{SDSS}
{York} D.~G. {et~al.}, 2000, \aj, 120, 1579

\end{thebibliography}

	\label{lastpage}
\end{document}